\documentclass[10pt,journal,compsoc]{IEEEtran}
\usepackage{array}
\usepackage[caption=false,font=normalsize,labelfont=sf,textfont=sf]{subfig}
\usepackage{textcomp}
\usepackage{stfloats}
\usepackage{url}
\usepackage{verbatim}
\usepackage{graphicx}
\usepackage{cite}
\usepackage{amsfonts}
\usepackage{amsmath}
\usepackage{amssymb}
\usepackage{mdframed}
\usepackage{mathtools}
\usepackage{tabularx}
\usepackage{pgfplots}
\usepackage{adjustbox}
\usepackage{paralist}
\usepackage{cleveref}
\usepackage{color}
\usepackage{svg}

\usepackage{algpseudocode}
\usepackage[ruled,linesnumbered,vlined]{algorithm2e}
\usepackage{booktabs}
\usepackage{enumitem}
\usepackage{threeparttable}
\usepackage{ragged2e}
\usepackage{multicol}
\usepackage{multirow} 
\usepackage{amsthm} 
\usepackage{balance} 

\usepackage{pifont}       
\usepackage{bbding}       
\usepackage{fontawesome}  

\newtheorem{theorem}{Theorem}

\newtheorem{definition}{Definition}

\usepackage{tikz}

\usepackage[table]{xcolor} 
\usepackage{colortbl}
\usepackage{wasysym} 

\begin{document}

\title{ShadowBlock: Efficient Dynamic Anonymous Blocklisting and Its Cross-chain Application}

\author{Haotian Deng,~\IEEEmembership{Student Member,~IEEE,} Mengxuan Liu,~\IEEEmembership{Student Member,~IEEE,} \\ Chuan Zhang,~\IEEEmembership{Member,~IEEE,} Wei Huang,~\IEEEmembership{Member,~IEEE,} \\ Licheng Wang,~\IEEEmembership{Member,~IEEE,} and Liehuang Zhu,~\IEEEmembership{Senior Member,~IEEE}

\thanks{This work is supported by the National Natural Science Foundation of China (Grant No. 62232002) and the Guangxi Science and Technology Program (Grant No. 2026GXNSFDA00640010).}

\IEEEcompsocitemizethanks{
    \IEEEcompsocthanksitem Haotian Deng is with the China United Network Communications Group Corporation Limited, Beijing, 100033, China, and the School of Cyberspace Science and Technology, Beijing Institute of Technology, Beijing 100081, China (e-mail: hdeng@bit.edu.cn). 
    
    \IEEEcompsocthanksitem Mengxuan Liu, Chuan Zhang, Wei Huang, Licheng Wang, and Liehuang Zhu are with the School of Cyberspace Science and Technology, Beijing Institute of Technology, Beijing 100081, China (e-mail: mengxuanliu@bit.edu.cn; chuanz@bit.edu.cn; huangwei@bit.edu.cn; lcwang@bit.edu.cn; liehuangz@bit.edu.cn).

   (Corresponding author: Chuan Zhang.)
   \vspace{5pt}
   }

}

\markboth{}%
{}

\IEEEpubid{}

\IEEEtitleabstractindextext{%
\begin{abstract}
Online harassment, incitement to violence, racist behavior, and other harmful content on social media can damage social harmony and even break the law. Traditional blocklisting technologies can block malicious users, but this comes at the expense of identity privacy. The anonymous blocklisting has emerged as an effective mechanism to restrict the abuse of freedom of speech while protecting user identity privacy. However, the state-of-the-art anonymous blocklisting schemes suffer from either poor dynamism or low efficiency, {requiring users to regenerate proofs after blocklist updates, incurring proof or verification overhead that grows with the blocklist size, or supporting only append-only updates without true removal.} In this paper, we propose $\mathsf{ShadowBlock}$, an efficient dynamic anonymous blocklisting scheme. Specifically, we utilize the pseudorandom function and cryptographic accumulator to construct the public blocklisting, enabling users to prove they are not on the blocklisting in an anonymous manner. To improve verification efficiency, we design an aggregation zero-knowledge proof mechanism that converts multiple verification operations into a single one. In addition, we leverage the accumulator's property to achieve efficient updates of the blocklisting, i.e., the original proof can be reused with minimal updates rather than regenerating the entire proof. Experiments show that $\mathsf{ShadowBlock}$ has better dynamics and efficiency than the existing schemes. Finally, the discussion on applications indicates that $\mathsf{ShadowBlock}$ also holds significant value and has broad prospects in emerging fields such as cross-chain identity management.
\end{abstract}

\begin{IEEEkeywords}
Anonymous blocklisting, identity authentication, non-membership proof, accumulator, zero-knowledge proof
\end{IEEEkeywords}}

\maketitle

\section{Introduction}

\IEEEPARstart{O}{nline} harms—including harassment, phishing, hate speech, incitement to violence, racist behavior, and fraud—pose severe negative consequences for society. For instance, Nasdaq’s Global Financial Crime Report\footnote{https://www.nasdaq.com/global-financial-crime-report} reveals that global fraud losses exceeded \$485 billion in 2023. Additionally, the Network Contagion Research Institute at Rutgers University\footnote{https://www.ohchr.org/en/statements/2023/01/freedom-speech-not-freedom-spread-racial-hatred-social-media-un-experts} noted that within 12 hours of Twitter’s acquisition, the use of hate speech and the racial slur ``N-word'' on the platform surged by nearly 500\% compared to the pre-acquisition average. Meanwhile, freedom of speech is a fundamental human right enshrined in international frameworks such as the Universal Declaration of Human Rights, and remains indispensable. It must, however, be exercised within the boundaries of laws designed to protect other fundamental rights. Consequently, balancing freedom of speech with social harmony and public security has become a pressing global challenge.

Blocklisting \cite{jung2001dns,ramachandran2007filtering,ramachandran2006revealing,xie2008spamming,kanich2008spamalytics,jung2002dns,sinha2010improving,google_safebrowsing_api,whittaker2010large,43,44,zhang4}, serves as an effective mechanism to restrict malicious behavior and abuse of freedom of speech, maintain social harmony and personal safety, and prevent the freedom of speech from being used to propagate hatred, spread rumors, or pose threats to public safety, thus safeguarding freedom of speech. {However, a significant challenge arises when banning an account on a platform \cite{ma2009learning,bell2020analysis,wiki_revocation,tls_revocations_2024,victor2025_clubcards,revdns2025,zhang5}. Users can easily create new accounts and continue to post freely.} Consequently, moderation tends to rely on robust centralized identity providers (e.g., Meta, Google, Apple) and linking social media platform identities to unique identifiers, like Social Security Numbers (SSN) in the United States, within and across platforms. Binding users' online speech to their real-world identities raises significant concerns regarding the Web’s decentralized ethos and user privacy, potentially resulting in a chilling effect on freedom of speech. Hence, the anonymous blocklisting came into being, which balances the needs of privacy and regulation, and enables blocking malicious users who violate platform rules while hiding their real identities.

Consequently, anonymous blocklisting technologies~\cite{33,34,35,PEREA,blac,SNARKBlock,SnarkFold,zk-Promises,fauzi,MaskAuct,quantum} have attracted considerable attention from both academia and industry. BLAC~\cite{blac} pioneered anonymous blocklisting in a fully trustless setting, allowing users to establish long-term pseudonymous identities in zero-knowledge and prove non-inclusion in the blocklist. However, it incurs substantial computational overhead on the client side. To address this inefficiency, SNARKBlock~\cite{SNARKBlock} introduced server-aided verification with only logarithmic complexity in blocklist size, dramatically reducing prover costs. Nevertheless, SNARKBlock reveals critical shortcomings in dynamic scenarios: removing even a single entry forces regeneration of non-membership proofs for all remaining users, resulting in quadratic time complexity relative to blocklist size. This fatal scalability bottleneck makes it impractical for real-world systems—particularly social media platforms that require real-time responsiveness, low latency, and frequent blocklist updates amid continuously expanding moderation targets. In a different application direction, MaskAuct~\cite{MaskAuct} proposed a specialized blocklisting mechanism for anonymous auctions, empowering sellers to autonomously exclude misbehaving bidders through a confidential blocklist while fully preserving bidder anonymity and bid privacy. Meanwhile, several general-purpose frameworks have also emerged. For instance, {zk-promises}~\cite{zk-Promises} provides a zero-knowledge verifiable callback mechanism on bulletin boards, but it suffers from high client-side overhead and poor dynamic scalability. Subsequently, ALPACA~\cite{ALPACA}, built on Incrementally Verifiable Computation (IVC)~\cite{ivc1,ivc2}, achieved asymptotically optimal constant-cost non-membership proofs. Despite its theoretical elegance and relatively heavy proving time with complex synchronization requirements, ALPACA’s blocklist remains strictly append-only: true revocation is impossible. It can only achieve “logical skipping” of lifted bans via ``ban-over signatures'', leaving historical records permanently on-chain. As a result, in long-running systems or scenarios with frequent unblocking, ALPACA inevitably accumulates massive redundant entries, significantly undermining its long-term practical efficiency. Although existing schemes have made progressive improvements in efficiency, specialization, or asymptotic performance, none yet simultaneously satisfy the core requirements of low client overhead, efficient dynamic updates, true revocation capability, and long-term scalability—leaving substantial room for further research and more practical solutions.

In this paper, we propose $\mathsf{ShadowBlock}$, an efficient dynamic anonymous blocklist scheme. Specifically, we utilize the pseudorandom function and cryptographic accumulator to construct the public blocklist, enabling users to prove they are not on the blocklist in an anonymous manner. To improve verification efficiency, we design an aggregation zero-knowledge proof mechanism to convert multiple verification operations into one. In addition, we design an accumulator-based membership proof mechanism to enable efficient updates of the blocklist, i.e., the original proof can be reused with minimal updates rather than regenerating the entire proof.
 
Our contributions are summarized as follows.

\begin{itemize}
    \item We introduce $\mathsf{ShadowBlock}$, which leverages pseudorandom functions and cryptographic accumulators to create a public blocklist that enables users to anonymously prove their non-inclusion.

    \item To enhance verification efficiency, an aggregation zero-knowledge proof mechanism is designed, condensing multiple verification operations into a single instance.

    \item An accumulator-based approach is proposed for updating the blocklist, enabling the reuse of the original proof with minimal updates, rather than re-generating the entire proof.

    \item We conduct extensive experiments and analysis on $\mathsf{ShadowBlock}$ to evaluate its performance and effectiveness. Additionally, we compared our scheme comprehensively with state-of-the-art works to demonstrate its superiority. 

    \item We design an application of $\mathsf{ShadowBlock}$, built upon decentralized identity (DID) \cite{w3c-DID,w3c-VC} technology, for cross-chain decentralized identity management, thereby broadening the applicability of blocklisting technology.

\end{itemize}

\section{Related Work}
\label{sec:2}
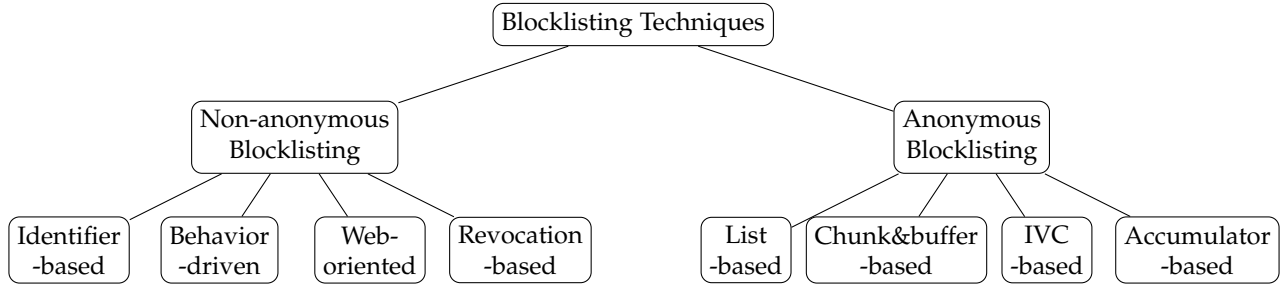
\begin{figure*}[t]
\centering
\begin{tikzpicture}[
    level 1/.style={sibling distance=90mm},
    level 2/.style={sibling distance=20mm},
    level 3/.style={sibling distance=10mm},
    every node/.style={draw, rectangle, rounded corners, align=center}
]

\node {Blocklisting Techniques}
  child {
    node {Non-anonymous\\Blocklisting}
      child { node {Identifier\\-based} }
      child { node {Behavior\\-driven} }
      child { node {Web-\\oriented} }
      child { node {Revocation\\-based} }
  }
  child {
    node {Anonymous\\Blocklisting}
      child { node {List\\-based} }
      child { node {Chunk\&buffer\\-based} }
      child { node {IVC\\-based} }
      child { node {Accumulator\\-based} }
  };

\end{tikzpicture}
\caption{A taxonomy of blocklisting techniques.
Non-anonymous approaches rely on linkable identifiers or globally visible states,
while anonymous blocklisting enforces exclusion without revealing identities}
\label{fig:blocklisting-taxonomy}
\end{figure*}

We delve into the literature on non-anonymous and anonymous blocklisting, offering an overview of the related work in this domain as shown in Fig. \ref{fig:blocklisting-taxonomy}.

\begin{table*}[htbp]
\scriptsize
\begin{center}
  \caption{{Comparison with Existing Anonymous Blocklisting Schemes}}
  \label{ComparisonPriorWorks}
  \renewcommand\arraystretch{1.1}
  \setlength{\tabcolsep}{2mm}{
  \begin{threeparttable}
  \begin{tabular}{|c|c|c|c|c|c|c|c|}
    \hline
    \centering
    {\textbf{Scheme}}
     & \textbf{{Category}}
     & \begin{tabular}[c]{@{}l@{}}\textbf{Misauthentication Resistance}\end{tabular} 
     & \textbf{Unlinkability} 
     & \textbf{Addition} 
     & \textbf{Removal} 
      &  \textbf{User Efficient}
       & \begin{tabular}[c]{@{}l@{}}\textbf{Verifier Efficient} \end{tabular}\\ \hline

    BLAC \cite{blac}
    & List-based
    & \CIRCLE 
    & \CIRCLE  
    & \Circle 
    & \Circle 
    & \Circle 
    & \Circle  \\ \hline 

    PEREA \cite{PEREA} 
    & List-based
    &  \CIRCLE 
    &  \CIRCLE  
    &  \CIRCLE 
    & \Circle
    & \Circle
    & \Circle\\ \hline

    MaskAuct \cite{MaskAuct}
    & List-based
    & \CIRCLE 
    & \CIRCLE  
    & \Circle 
    & \Circle 
    & \Circle 
    & \Circle  \\ \hline

    SNARKBlock \cite{SNARKBlock} 
    & Chunk\&buffer-based
    &  \CIRCLE 
    &  \CIRCLE 
    &  \CIRCLE 
    & \Circle
    & \RIGHTcircle 
    & \CIRCLE\\ \hline

    {zk-promises} \cite{zk-Promises} 
    & Chunk\&buffer-based
    &  \CIRCLE 
    &  \CIRCLE 
    &  \CIRCLE
    & \RIGHTcircle  
    & \RIGHTcircle 
    & \CIRCLE\\ \hline

    ALPACA \cite{ALPACA}
    & IVC-based
    &  \CIRCLE 
    &  \CIRCLE 
    & \CIRCLE 
    & \CIRCLE 
    & \RIGHTcircle
    & \CIRCLE \\ \hline 

    \rowcolor[HTML]{C0C0C0} $\mathsf{ShadowBlock}$
    & Accumulator-based
    &  \CIRCLE 
    & \CIRCLE 
    & \CIRCLE 
    & \CIRCLE  
    & \CIRCLE
    & \CIRCLE \\ \hline
  \end{tabular}

  \begin{tablenotes}
        \footnotesize
       \item \CIRCLE Support;  \Circle Not support;  \RIGHTcircle Limited support 
  \end{tablenotes}
  \end{threeparttable}}
  \end{center}
  \vspace{-0.2in}
\end{table*}

\subsection{Non-anonymous Blocklisting}

\textbf{Identifier-based Blocklisting.}
{Early blocklisting mechanisms rely on explicit and stable identifiers, such as user accounts, IP addresses, or domain names, and reduce enforcement to membership tests over a maintained deny list. These systems are widely deployed in spam filtering and network security due to their simplicity and low latency \cite{jung2001dns,ramachandran2007filtering}. However, their effectiveness fundamentally depends on identifier persistence. In adversarial settings, attackers can evade enforcement through identity churn, such as dynamic IP allocation, botnets, or mass account registration, forcing frequent updates and increasing false positives \cite{ramachandran2006revealing,xie2008spamming}. Empirical studies show that IP-based blocklists are ineffective against large-scale automated abuse, as botnet diversity enables rapid circumvention \cite{ramachandran2007filtering,kanich2008spamalytics}. These approaches assume stable and scarce identifiers, which does not hold in adversarial environments.}

\textbf{Behavior-driven Blocklisting.}
{To address identifier evasiveness, behavior-driven approaches derive blocklisting decisions from observable activity patterns, such as sending rates and traffic correlations, rather than static identities. This shifts blocklisting from set membership to statistical inference over aggregated observations, improving robustness against identity rotation. However, such systems must integrate noisy data from distributed sources, making updates sensitive to threshold selection and policy inconsistency. Sinha et al.\ highlight this instability and propose dynamic thresholding with speculative aggregation to balance responsiveness and false positives \cite{sinha2010improving}. These methods still rely on linkable observations or persistent attributes, and introduce system-level complexity in maintaining consistent policies.}

\textbf{Web-oriented Blocklisting.}
{Web-oriented blocklisting targets large and rapidly evolving item sets, such as URLs and domains, requiring scalable distribution and low-latency client-side enforcement. Systems such as Google Safe Browsing adopt hash-prefix-based lookup to reduce bandwidth while enabling frequent updates \cite{google_safebrowsing_api}. On the detection side, machine learning techniques have been widely used to identify phishing content at scale \cite{whittaker2010large,ma2009learning}. Measurement studies further show that different blocklists vary significantly in coverage, freshness, and accuracy, motivating multi-source aggregation and hybrid enforcement strategies \cite{bell2020analysis}. These systems prioritize scalability and responsiveness, but operate on globally visible items, making user interactions and enforcement inherently linkable.}

\textbf{Revocation-based Blocklisting:}
{Revocation-based approaches emphasize correctness and auditability using authenticated data structures or signed responses. In PKI systems, Certificate Revocation Lists (CRLs) and OCSP provide verifiable revocation status \cite{wiki_revocation}. However, CRLs incur significant bandwidth and storage overhead, while OCSP introduces latency and potential privacy leakage. Empirical studies show that achieving complete and timely revocation coverage remains challenging \cite{tls_revocations_2024}. Practical systems such as CRLite and CRLSet improve efficiency through compression and centralized distribution, but still face trade-offs between freshness, scalability, and auditability \cite{victor2025_clubcards}. Emerging decentralized approaches further introduce consistency and deployment complexity \cite{revdns2025}. These methods rely on globally visible revocation states and introduce trade-offs between scalability, freshness, and privacy.}

\textbf{Summary.}
{Non-anonymous blocklisting techniques are mature and widely deployed, emphasizing efficiency and scalability across different domains. However, they fundamentally rely on explicit or linkable identities and globally observable states, making user behavior and enforcement decisions traceable. This limitation renders them unsuitable for privacy-preserving scenarios and motivates the need for anonymous blocklisting mechanisms.}

\subsection{Anonymous Blocklisting}

The anonymous blocklisting has received widespread attention. There have been many related works \cite{25,26,27,28,29,30,31,32} in both the academic and industrial fields\cite{13,14,15,16,17,18,zhang6}. We classify anonymous blocklisting into four categories, as shown in Table \ref{ComparisonPriorWorks}. 

{\textbf{List-based.} Early anonymous blocklisting systems \cite{19,20,21,22,23,24,blac,PEREA} such as BLAC \cite{blac} and PEREA \cite{PEREA} laid the foundational groundwork for enforcing accountability in privacy-preserving environments. BLAC \cite{blac} was the first to support long-term anonymous identities and zero-knowledge non-membership proofs in a trustless setting, but its computational and communication overheads made practical deployment difficult. PEREA \cite{PEREA} improved upon BLAC \cite{blac} by using token generation to avoid traversing the entire blocklisting, thereby reducing the verification burden; however, its fixed sliding-window mechanism still limited scalability. MaskAuct \cite{MaskAuct} offers anonymous blocklisting tailored for sealed-bid auction systems, allowing sellers to exclude misbehaving bidders without compromising bid confidentiality. However, such systems are not designed for general-purpose, dynamically evolving blocklists.}

{\textbf{Chunk\&buffer-based}. SNARKBlock \cite{SNARKBlock}, leveraged chunk\&buffer blocklist structure to achieve logarithmic-time server-side verification, significantly reducing verification costs. Yet, because SNARKBlock’s proof structure depends on the global consistency of the entire blocklist, deleting even a single entry requires every user to regenerate their non-membership proofs, resulting in quadratic overhead in dynamic settings. This bottleneck, combined with the substantial on-chain storage required for storing blocklist tuples, makes SNARKBlock increasingly impractical, particularly in blockchain-based environments where storage is scarce and costly. In contrast, zk-promises \cite{zk-Promises} provides a fully general zero-knowledge framework supporting turing-complete anonymous state machines and verifiable callbacks, enabling sophisticated moderation logic such as anonymous reputation, rate limiting, and asynchronous penalties. Nonetheless, {zk-promises} \cite{zk-Promises} incurs substantial client overhead, offers only partial support for efficient removal, and scales poorly under frequent updates.}

{\textbf{VC-based.} ALPACA \cite{ALPACA}, built on incrementally verifiable computation (IVC), achieves constant-size proofs and constant-time updates, earning strong marks in user and verifier efficiency. Yet its append-only blocklist structure allows only logical overrides rather than true removal, causing historical entries to accumulate indefinitely.}

{\textbf{Accumulator-based.} Against this backdrop, combining strong anonymity guarantees with full support for both additions and true removals, $\mathsf{ShadowBlock}$ uses dynamic accumulators to remove blocklist entries at the mathematical level—something neither SNARKBlock nor ALPACA can achieve. Moreover, its use of proof aggregation and efficient accumulator updates dramatically reduces client and verifier overhead. As a result, $\mathsf{ShadowBlock}$ is one of the few systems in the comparison table that achieves simultaneously robust dynamic flexibility, low overhead, and strong scalability, making it particularly well suited for real-world, high-throughput environments such as social media platforms and decentralized identity systems that require rapid and continuous moderation actions.}

\textbf{Summary.} $\mathsf{ShadowBlock}$ uniquely combines correctness, anonymity, true dynamic update capability, and practical efficiency, giving it the strongest overall profile among existing anonymous blocklisting schemes.

\section{Preliminaries}
\label{sec:3}
In this section, we review some background knowledge used
in our work, pseudorandom function \cite{PRF}, cryptographic accumulator \cite{accumlator} and
zero-knowledge proof \cite{zkp}.

\subsection{Pseudorandom Function}

A pseudorandom function (PRF) \cite{PRF} is a fundamental component in constructing order-revealing encryption schemes. Let $\mathcal{K}$ denote the key space, $\mathcal{X}$ the input domain, and $\mathcal{Y}$ the output range. We define a function $\mathsf{PRF}: \mathcal{K} \times \mathcal{X} \rightarrow \mathcal{Y}$ to be a secure pseudorandom function if, for a randomly selected key $\mathsf{k} \in \mathcal{K}$, no adversary can distinguish the outputs of $\mathsf{PRF}_k( \cdot)$ from those of a truly random function $f: \mathcal{X} \rightarrow \mathcal{Y}$ with any significant advantage.

\subsection{Cryptographic Accumulator}

A general accumulator \cite{accumlator}, in the context of cryptography, is defined as a data structure that enables efficient addition of elements and subsequent verification of whether a specific element belongs to the accumulated set. It is designed to provide membership proofs without disclosing the individual elements stored within the accumulator.

Formally, a general accumulator consists of the following components:

\begin{itemize}
    \item $\mathsf{Setup}$: A setup algorithm that generates the initial parameters and state required for the accumulator.
     \item $\mathsf{Add}$: An add algorithm that takes an element as input and efficiently incorporates it into the accumulator.
    \item $\mathsf{Verify}$: A verification algorithm that takes an element and a proof, determining whether the element is a member of the accumulated set.
    \item $\mathsf{Non-membership\ Proof}$: A non-membership proof algorithm that proves that an element is not a member of the accumulated set.  
\end{itemize}

The general accumulator has the following properties:
\begin{itemize}
    \item \textbf{Correctness:} The verification algorithm should accurately judge the membership of elements in the accumulated set.
    \item \textbf{Efficiency:} The add and verify algorithms should be computationally efficient, allowing for fast addition and verification operations.
    \item \textbf{Soundness:} The accumulator should be secure against unauthorized modifications or tampering, ensuring the integrity of the accumulated set.
    \item \textbf{Privacy: }The accumulator should protect the privacy of individual elements, preventing inference or reconstruction of the original set from the accumulator.
\end{itemize}

The basic security property for accumulators is soundness (sometimes called undeniability \cite{33}) which states that an adversary cannot construct an accumulator $A$ and a set $I$ for which both $\pi_I$ and $\Bar{\pi_I}$ are simultaneously valid. Below we state strong soundness (also found in \cite{10}), which allows the adversary to create the accumulator without revealing the accumulated set $X$ to the challenger.

\subsection{Zero-Knowledge Proof}

Zero-knowledge proof (ZKP) \cite{zkp} technology was proposed by Goldwasser et al. \cite{zk}, which is a two-party cryptographic protocol running between a prover and a verifier that can be used to perform knowledge proofs. Specifically, they offer a way for one party, called the prover, to convince another party, called the verifier, that a certain statement is true without revealing any additional information beyond its truth. Zero-knowledge proofs have three important properties:

\begin{itemize}
    \item Completeness: If the statement being proven is true, an honest verifier will be convinced of its truth by an honest prover. In other words, if the prover knows the secret information, the verifier will accept the proof.
    \item Soundness: If the statement being proven is false, no prover, even a dishonest one, can convince the verifier otherwise except with negligible probability. This property ensures a dishonest prover cannot trick the verifier into accepting a false statement.
    \item Zero-knowledge: The proof reveals nothing about the secret information or any additional information that could help the verifier deduce it. The verifier only gains information on whether the statement is true or false.
\end{itemize}

Zero-knowledge proofs have found applications in various fields, including secure authentication, secure computation, and privacy-preserving protocols. They provide a powerful tool for demonstrating knowledge or possessing private information without compromising privacy or revealing sensitive data.

\subsubsection{Groth16}
We briefly describe the trusted-setup zero-knowledge succinct non-interactive argument of knowledge (zkSNARK) scheme \cite{Groth16}. At a high level, given a description of an arithmetic circuit (over the scalar field of a pairing-friendly elliptic curve), a Groth16 proof proves that a circuit is satisfied by a set of public wires (values known to the verifier) and private wires (values which are not known to the verifier, also called witness elements).

The main advantage of Groth16 lies in its efficiency and scalability. It achieves this by utilizing a technique called the ``zk-SNARK'' construction. This construction makes proof sizes significantly smaller than previous systems, making it practical for various applications. Groth16 consists of three main steps:

\begin{itemize}
    \item $\mathsf{ZKSetup}(1^\lambda \rightarrow pp)$: Given a security parameter $\lambda$, the algorithm compiles the arithmetic circuit into R1CS constraints, generates public parameters $pp$ through a trusted setup with elliptic curve operations, and outputs $pp$ as the global cryptographic setup.
     \item $\mathsf{ZKProve}(pp, x, w \rightarrow \pi)$: Leveraging public parameters $pp$, the prover first computes a witness $w$ satisfying the Rank-1 Constraint System (R1CS) for public input $x$, then transforms the R1CS system into a Quadratic Arithmetic Program (QAP) to construct polynomial commitments, and ultimately generates a succinct zero-knowledge proof $\pi$ composed of group elements over elliptic curve groups.
    \item $\mathsf{ZKVerify}(pp, x, \pi \rightarrow {0,1})$: The verifier checks the $\pi$ with $pp$ and public input $x$, returning 1 if the proof $\pi$ is valid, otherwise 0.
\end{itemize}

\section{System Overview}
\label{sec:4}

This section briefly introduces $\mathsf{ShadowBlock}$ from three aspects: {system model, design goals and threat model.}

\subsection{System Model}

\begin{figure}[htpb!]
	\centering
	\includegraphics[width=0.45\textwidth]{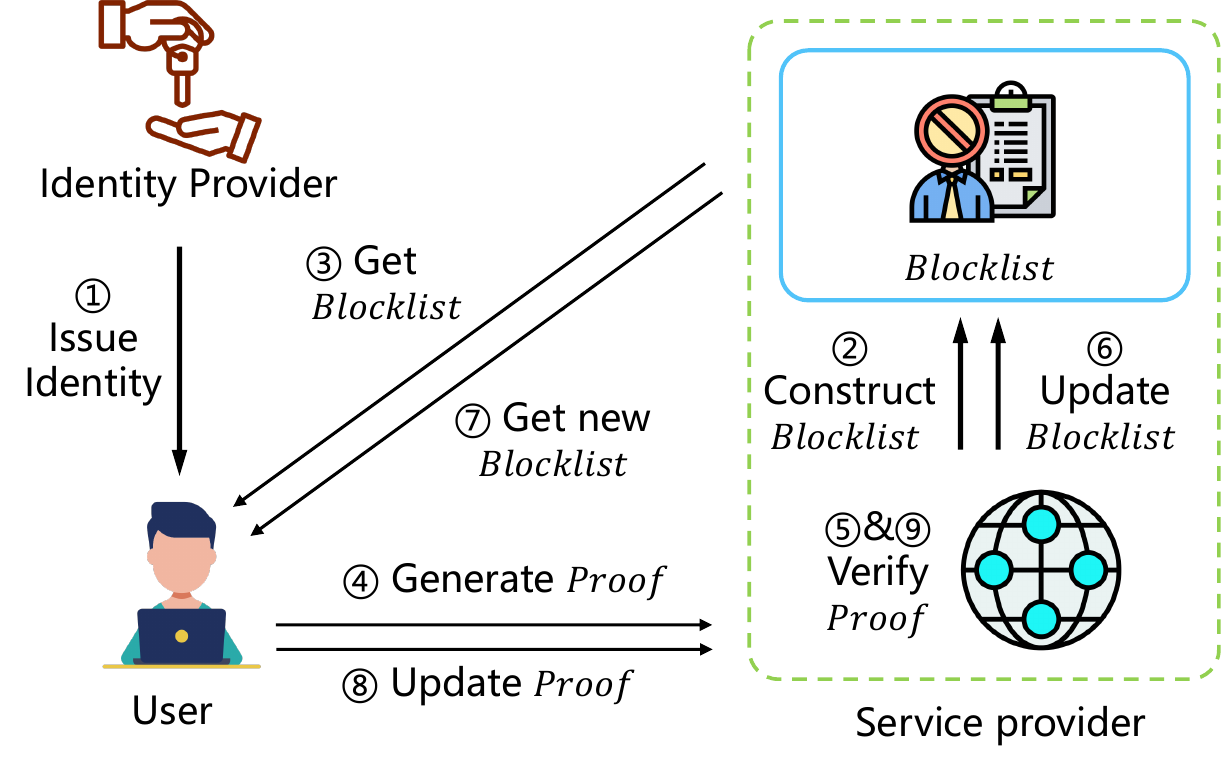}
	\caption{System model}\label{system_model}
\end{figure}

As shown in Fig. \ref{system_model}, $\mathsf{ShadowBlock}$ consists of three main entities: Identity Provider, User, and Service Provider.

\begin{itemize}
    \item \textbf{Identity provider.} The identity provider issues a user with an identity (Step $\mathbf{1}$) that can be recognized by multiple service providers and assigns a unique identity key (private key) to the user for interaction with the service provider. Notably, identity providers are mostly major internet corporations, including Google Workspace from Google, Azure Active Directory by Microsoft, and Amazon's Identity and Access Management, etc.

    \item \textbf{User.} The user must pass identity authentication to use the service provider's services. First, the user needs to obtain the blocklist from the service provider (Step $\mathbf{3}$), and then generate a proof based on the blocklist information (Step $\mathbf{4}$) to be authenticated by the service provider. In addition, when the blocklist is dynamically updated, the user needs to obtain a new blocklist (Step $\mathbf{7}$) and directly update the original proof for the new blocklist (Step $\mathbf{8}$), instead of recalculating the entire proof. The user sends the updated proof to the service provider for verification.

    \item \textbf{Service provider.} The service provider provides users with services in the digital world, but the service provider does not hold the user's identity information, and users are anonymous to the service provider (Step $\mathbf{2}$). The service provider is expected to build a blocklist list, and then it verifies the proof sent by the user (Step $\mathbf{5}$). If the verification passes, the user is allowed to use the service; if not, the service is refused. In addition, if the service provider finds that the user uses the service in violation of laws and regulations, or the user appeals or the ban time limit has passed, the service provider shall immediately update the blocklist (Step $\mathbf{6}$). After the blocklist is updated, the service provider only verifies the proof based on the latest blocklist(Step $\mathbf{9}$). The service provider tends to be social media, such as Twitter, WeChat, Instagram, etc.

\end{itemize}

 Specifically, $\mathsf{ShadowBlock}$ is divided into blocklist authentication and management. The blocklist authentication employs pseudorandom functions and zero-knowledge proofs to enable users to prove they are not on a blocklist without revealing their real identities, and leverages the accumulator to significantly reduce the storage size of the blocklist. The blocklist management employs RSA dynamic accumulators for efficient addition and removal of members, ensuring real-time updates and aggregation techniques consolidate multiple zero-knowledge proofs (ZKPs), reducing computational and verification overhead and enhancing scalability and privacy in large-scale systems.

\subsection{Design Goals}

The design of $\mathsf{ShadowBlock}$ is driven by six objectives: blockability, unforgeability, anonymity, dynamic, compact storage and scalable verification.

{\textbf{{Blocklistability.}} The user can only successfully authenticate to a service provider if that user holds a valid key issued by an identity provider that is not included in the blocklist.}

{\textbf{{Unforgeability.}} An adversary $\mathcal{A}$ cannot forge and prevent another unblocked user from successfully authenticating with a service provider.}

{\textbf{{Anonymity.}} The service providers cannot distinguish authentication proofs associated with any two users. Furthermore, no such coalition can link any authentication proof with the registration in which an identity is registered. $\mathsf{ShadowBlock}$ derives session-level pseudonymous tags from a secret key and fresh randomness
using a pseudorandom function. As a result, the same user
appears unlinkable across different authentication sessions.}

{\textbf{{Compact storage.}}
Maintaining the entire blocklist explicitly is inefficient in large-scale systems, especially when public verifiers must repeatedly check non-membership. $\mathsf{ShadowBlock}$ therefore uses a dynamic RSA accumulator
to compress the blocklist into a constant-size public value,
enabling efficient representation and verification.}

{\textbf{{Scalable verification.}}
In high-throughput scenarios, verifying many independent proofs may become a bottleneck. $\mathsf{ShadowBlock}$ introduces proof aggregation to combine multiple proofs into a single verification object, thereby reducing the overall verification overhead while preserving privacy guarantees.}

{\textbf{{Dynamic.}} $\mathsf{ShadowBlock}$ is to enable a blocklist that can be efficiently modified or updated over time. Such a dynamic blocklist should not only adjust to changes when necessary but also allow users to incorporate these updates into their existing authentication proofs, eliminating the need to recreate the entire proof from scratch.}

\subsection{Threat Model}

$\mathsf{ShadowBlock}$ follows an asynchronous communication model that allows messages that are not delivered, delayed, or erroneous to exist. The network environment may be subject to passive and active attacks. Specifically, active attackers may attempt to modify messages or disrupt communication between users, identity providers, or service providers. Passive attackers may attempt to eavesdrop on messages or analyze user traffic to obtain users' private information. We specifically give the adversary abilities to attack $\mathsf{ShadowBlock}$:

{\textbf{Cryptographic adversary.}}
\begin{itemize}
    \item {An adversary $\mathcal{A}$ can eavesdrop, forge, modify, delay, or remove messages between any two communicated entities.}
     \item {The service provider is a semi-honest (i.e., honest but curious) entity that provides reliable services, but it will try to learn the private information about users' real identities. }
     \item The adversary $\mathcal{A}$ can access the blocklist and
get all the information stored on the blocklist, but $\mathcal{A}$ cannot modify the information of the blocklist.
    \item {All entities of the system are the honest majority.}
\end{itemize}

{\textbf{System-level adversary.}}

In addition to cryptographic attacks, we consider adversaries exploiting the cross-chain and distributed deployment environment:

\begin{itemize}
    \item {Due to propagation delays between the relay chain and target chains, an adversary may attempt to use an outdated accumulator state and its corresponding witness.}
    \item {Different chains may temporarily maintain inconsistent views of the blocklist state.} 
\end{itemize}

\section{Blocklist Authentication}
\label{sec:5}

Blocklist authentication leverages an RSA accumulator to represent the blocklist compactly while preserving user anonymity. 
Each user derives a pseudorandom identifier $\mathsf{tag}$ from its secret identity key $k$ and a fresh session value $\mathsf{nonce}$. 
When authenticating to the service provider, the user proves in zero knowledge that its identifier is not accumulated in the current blocklist, without revealing either $k$ or $\mathsf{tag}$ itself. 
This section presents the initialization, blocklist construction, and non-membership proof generation procedures. {We summarize the main notations used in the following sections as shown in Table \ref{tab:notation}.}

\begin{table}[h]
\centering
\caption{{Symbol \& Description}}
\label{tab:notation}
\begin{tabular}{@{}ll@{}}
\toprule
\textbf{Symbol} & \textbf{Definition} \\
\midrule
$k$ & User's secret identity key issued by the identity provider \\
$\mathsf{nonce}$ & Fresh randomness associated with a session \\
$\mathsf{tag}$ & Pseudorandom anonymous identifier \\
$X$ & Set of banned pseudorandom identifiers \\
$Y$ & Set of corresponding hash values \\
$\pi$ & Zero-knowledge non-membership proof \\
${\mathcal{A}(Y)}$ & Accumulator value representing the blocklist \\
${{\mathcal{A}'(Y)}}$ & Accumulator after addition \\
${\mathcal{A}''(Y)}$ & Accumulator after removal \\
$C_{\mathsf{agg}}$ & Aggregated commitment \\
$\pi_{\mathsf{agg}}$ & Aggregated zero-knowledge proof \\
\bottomrule
\end{tabular}
\end{table}

\subsection{Initialization}

The service provider initializes the blocklist system by generating the public parameters for the RSA accumulator, the zero-knowledge proof system, and the pseudorandom function.

\subsubsection{RSA Parameter Generation}

The service provider selects two large distinct primes $p$ and $q$, and computes the RSA modulus $N=pq$. 
It then chooses a generator $g \in \mathbb{Z}_N^{*}$ for the accumulator domain. 
Optionally, a public exponent $e$ and its inverse $d \bmod \varphi(N)$ may also be generated for auxiliary RSA-based operations. 
The accumulator itself is defined over $(N,g)$.

\subsubsection{Zero-Knowledge Proof System Setup}

The service provider initializes a zero-knowledge proof system, such as zk-SNARK~\cite{zkSnark}, and runs the setup algorithm to obtain public parameters
\[
pp=\mathsf{ZKSetup}(1^\lambda),
\]
where $\lambda$ denotes the security parameter. 
The public parameters $pp$ are then published for use in subsequent proof generation and verification.

\subsubsection{PRF Configuration}

The service provider specifies a secure pseudorandom function $\mathsf{PRF}$ and distributes its description to all users. 
Each user holds a private identity key $k$ issued by the identity provider and uses it to derive pseudorandom anonymous identifiers.

\subsection{Blocklist Construction}

After initialization, the service provider monitors user behavior and determines whether a user should be blocked. 
If a user violates platform policies, the corresponding identifier is inserted into the blocklist and the accumulator is updated accordingly.

\subsubsection{Blocklist Structure}

\begin{figure}[htpb!]
	\centering
	\includegraphics[width=0.45\textwidth]{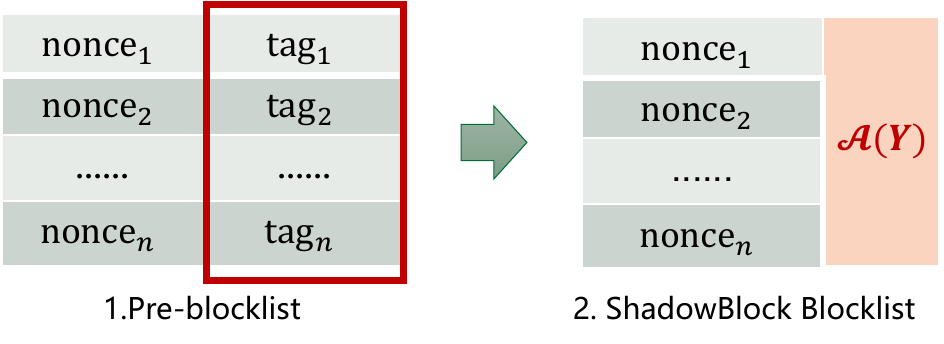}
	\caption{{Blocklist structure}}
	\label{blocklist}
\end{figure}

Before compression, the blocklist may be represented as
\[
L_{\mathrm{pre}}=\{(\mathsf{nonce}_1,\mathsf{tag}_1),\dots,(\mathsf{nonce}_n,\mathsf{tag}_n)\},
\]
where each $\mathsf{nonce}_i$ is associated with a session or message interaction, and $\mathsf{tag}_i$ is the corresponding anonymous identifier derived from the PRF.  As shown in Fig. \ref{blocklist}
However, storing all pairs explicitly incurs substantial storage overhead, especially in blockchain-based deployments where transparency and tamper-resistance are required but on-chain storage remains expensive~\cite{zhang2}.To reduce storage cost, $\mathsf{ShadowBlock}$ compresses the blocklist into an RSA accumulator. 
Let $X=\{\mathsf{tag}_1,\dots,\mathsf{tag}_n\}$
denote the set of banned pseudorandom identifiers, and let $Y=\{y_1,\dots,y_n\}$
be the set of corresponding hash values, where each $y_i = H_{\mathsf{prime}}(\mathsf{tag}_i)$
is obtained from a hash-to-prime function. 
The current blocklist is then represented by
\[
L=\{(\mathsf{nonce}_1,\dots,\mathsf{nonce}_n),{\mathcal{A}(Y)}\},
\]
where the accumulator value is defined as
\[
{\mathcal{A}(Y)}=g^{\prod_{y\in Y}y}\bmod N.
\]

Since the size of ${\mathcal{A}(Y)}$ is constant, the storage overhead of the compressed blocklist does not grow with the number of blocked identifiers. 
Therefore, the proposed structure is significantly more compact than storing all $\mathsf{tag}_i$ explicitly, while still allowing public and tamper-evident publication on the blockchain.

\subsubsection{Accumulator Generation}
\label{sec:5.2.2}

For a session with fresh randomness $\mathsf{nonce}'$, the user derives a pseudorandom anonymous identifier
\[
\mathsf{tag}'=\mathsf{PRF}_k(\mathsf{nonce}').
\]
If this identifier is determined to be blocklisted, the service provider computes its corresponding prime representative
\[
y' = H_{\mathsf{prime}}(\mathsf{tag}').
\]
The updated sets become
\[
X' = X \cup \{\mathsf{tag}'\}, \qquad Y' = Y \cup \{y'\}.
\]
The accumulator is then updated as
\[
{\mathcal{A}'(Y)}={\mathcal{A}(Y)}^{y'} \bmod N.
\]

Finally, the service provider publishes the updated blocklist
\[
L'=\{(\mathsf{nonce}_1,\dots,\mathsf{nonce}_n,\mathsf{nonce}'),{\mathcal{A}'(Y)}\}
\]
on the blockchain. 
This publication ensures tamper-evident transparency while preserving constant-size representation of the blocklist through the accumulator.

\subsection{Non-Membership Proof Generation}

To authenticate successfully, a user must prove that its identifier is not included in the current blocklist represented by ${\mathcal{A}(Y)}$. 
This proof should reveal neither the secret key $k$ nor the identifier $\mathsf{tag}$.

\subsubsection{Generate Zero-Knowledge Proof}

The user first computes
\[
\mathsf{tag}'=\mathsf{PRF}_k(\mathsf{nonce}'),
\qquad
y' = H_{\mathsf{prime}}(\mathsf{tag}').
\]

To prove that $\mathsf{tag}' \notin X$, the user constructs a standard RSA accumulator non-membership witness. 
Concretely, the user finds integers $(a,b)$ satisfying the Bézout relation
\[
a \cdot y' + b \cdot \prod_{y \in Y} y = 1.
\]
Using $a$, the user computes
\[
d = g^a \bmod N.
\]
The resulting witness satisfies the RSA non-membership equation
\[
d^{\,y'} \cdot {\mathcal{A}(Y)}^b \equiv g \pmod N.
\]

The user then generates a zero-knowledge proof
\[
\pi=\mathsf{ZKProve}\bigl(pp,\text{witness}=(k,\mathsf{nonce}',\mathsf{tag}',y',d,b)\bigr),
\]
proving the following statements simultaneously:

\begin{enumerate}
    \item $\mathsf{tag}'=\mathsf{PRF}_k(\mathsf{nonce}')$;
    \item $y'=H_{\mathsf{prime}}(\mathsf{tag}')$;
    \item $d^{\,y'} \cdot {\mathcal{A}(Y)}^b \equiv g \pmod N$.
\end{enumerate}

The user sends $(\mathsf{nonce}',\pi)$ to the service provider. 
Only one proof is generated for the user's own hidden identifier, and the proof size remains constant.

\subsubsection{Non-Membership Proof Verification}

Upon receiving $(\mathsf{nonce}',\pi)$, the service provider verifies the proof with public parameters $pp$ and public inputs $({\mathcal{A}(Y)},N,g,\mathsf{nonce}')$:
\[
\mathsf{Result}=\mathsf{ZKVerify}(pp,\pi,({\mathcal{A}(Y)},N,g,\mathsf{nonce}')).
\]

If $\mathsf{Result}=1$, the proof is accepted, meaning that the user has successfully proved in zero knowledge that its identifier is not included in the current blocklist. 
The service provider may then grant access. 
If $\mathsf{Result}=0$, the proof is rejected and access is denied.

This section establishes a privacy-preserving authentication mechanism using RSA accumulators, PRFs, and zero-knowledge proofs. 
Instead of revealing the anonymous identifier itself, the user only proves its non-membership in the accumulator. 
However, when the blocklist changes frequently, repeatedly regenerating such proofs becomes costly. 
The next section introduces dynamic witness updates and proof aggregation to address this issue efficiently.

\section{Blocklist Management}
\label{sec:6}

This section presents the dynamic management of the blocklist. 
The service provider may add newly offending users to the blocklist or remove users whose ban period has expired or whose appeals have been accepted. 
To support such updates efficiently, $\mathsf{ShadowBlock}$ updates the accumulator and allows users to refresh their non-membership proofs incrementally. 
In addition, multiple proofs can be aggregated to reduce batch verification overhead.

\subsection{Update Members from the Blocklist}

The blocklist evolves over time through addition and deletion operations. 
The addition procedure has already been described in Section~\ref{sec:5.2.2}. 
We now present the corresponding removal procedure and its effect on user proofs.

\subsubsection{Update the Accumulator}

Suppose a blocked identifier $\mathsf{tag}'' \in X$ is removed from the blocklist after an appeal or the expiration of the suspension period. 
Let
\[
y'' = H_{\mathsf{prime}}(\mathsf{tag}'').
\]
The service provider updates the blocklist sets as
\[
X \leftarrow X \setminus \{\mathsf{tag}''\},
\qquad
Y \leftarrow Y \setminus \{y''\}.
\]

The updated accumulator after removal is denoted by ${\mathcal{A}''(Y)}$ and corresponds to the new set $Y \setminus \{y''\}$:
\[
{\mathcal{A}''(Y)} = g^{\prod_{y \in Y \setminus \{y''\}} y}\bmod N.
\]

Finally, the service provider publishes the updated blocklist
\[
L''=\{(\mathsf{nonce}_1,\dots,\mathsf{nonce}_n),{\mathcal{A}''(Y)}\}
\]
on the blockchain, ensuring transparent and tamper-evident recording of the new accumulator state.

{In practical deployments, multiple moderators or administrators may issue blocklist updates concurrently. Since the accumulator state is globally shared, $\mathsf{shadowblock}$ assumes that accepted add/remove operations are first ordered by an upper-layer governance mechanism, such as an administration committee or sequencer, and are then applied to the accumulator in that order. Therefore, accumulator updates are logically serialized, while concurrency is resolved at the state-ordering layer rather than within the cryptographic primitive itself.}

\subsubsection{Updates to Non-Membership Proofs}

Once the blocklist is updated, previously generated non-membership witnesses may no longer correspond to the current accumulator state. 
A naive approach is to regenerate all proofs from scratch, but this incurs substantial computational overhead. 
To avoid this, $\mathsf{ShadowBlock}$ supports incremental witness updates.

Let the current blocklist at epoch $t$ be denoted by $\mathbb{B}_t$, and let the update operation be
\[
\delta_t:\mathbb{B}_t \rightarrow \mathbb{B}_{t+1}.
\]
If a new identifier is added, we write
\[
\delta_t = \delta_{\mathrm{add}}(\mathsf{nonce}',\mathsf{tag}'),
\]
whereas if an identifier is removed, we write
\[
\delta_t = \delta_{\mathrm{del}}(\mathsf{nonce}'',\mathsf{tag}'').
\]

Users retrieve the new accumulator value
\[
{A^{*}(Y)}=
\begin{cases}
{\mathcal{A}'(Y)}, & \delta_t=\delta_{\mathrm{add}},\\[4pt]
{\mathcal{A}''(Y)}, & \delta_t=\delta_{\mathrm{del}}.
\end{cases}
\]

Instead of regenerating the full proof immediately, a user first updates its non-membership witness relative to the new accumulator state, and then derives a refreshed proof
\[
\pi_{\mathrm{new}}=\mathsf{ZKProve}(pp,\text{updated witness}).
\]
The detailed witness update procedure is described below.

\subsection{Aggregation of Non-Membership Proofs}
\label{sec:6.2}

To improve efficiency in large-scale deployments, $\mathsf{ShadowBlock}$ aggregates multiple non-membership proofs into a single batch proof. 
This reduces verification overhead for the service provider and improves scalability when many users authenticate against the same blocklist state. All proofs in the batch correspond to the same epoch $\mathsf{epoch}$ and accumulator state ${\mathcal{A}'(Y)}$. Besides, We introduce an epoch parameter $\mathsf{epoch}$ to represent the current version of the blocklist. 
Each time the blocklist is updated (either by addition or removal), the service provider increments $\mathsf{epoch}$ and publishes the updated accumulator together with the new epoch value.

The epoch is included in proof generation and aggregation to ensure that all proofs are bound to the same blocklist state, and to prevent replay attacks across different blocklist versions.

\subsubsection{Aggregation Process}

{The user performs the following operational procedures. First, the user collects the complete set of non-membership proofs
\[
\pi_{\mathrm{all}}=\{\pi_1,\pi_2,\dots,\pi_m\},
\]
while obtaining the corresponding blocklist $L'$ containing the updated accumulator value ${\mathcal{A}'(Y)}$ and the nonce sequence $(\mathsf{nonce}_1,\mathsf{nonce}_2,\dots,\mathsf{nonce}_m)$.}

{Next, for each user $i$, the prover computes a commitment digest that binds the corresponding witness to the current accumulator state:}
\[
{
C_i = H\!\left(\mathsf{epoch}\,\|\,{\mathcal{A}'(Y)}\,\|\,d_i\,\|\,b_i\right).
}
\]

{These individual commitments are then aggregated into a single digest:}
\[
{
C_{\mathsf{agg}} = H(C_1\|\cdots\|C_m).
}
\]

{Based on the RSA accumulator properties, the user constructs an aggregated zero-knowledge proof $\pi_{\mathsf{agg}}$, which integrates multiple non-membership verification conditions into a unified proof framework through composite circuit constraints, satisfying:}
\[
{
\pi_{\mathsf{agg}}
=
\mathsf{ZKProve}_{\mathsf{agg}}
\left(
pp,\{\mathsf{witness}_i\}_{i=1}^{m},\mathsf{statement}_{\mathsf{agg}}
\right).
}
\]

{The aggregated statement is defined as:}
\[
{
\mathsf{statement}_{\mathsf{agg}}
=
\left(
\mathsf{epoch},\,
{\mathcal{A}'(Y)},\,
C_{\mathsf{agg}},\,
\mathsf{nonce}_1,\dots,\mathsf{nonce}_m
\right).
}
\]

{The aggregated witness is defined as:}
\[
{
\mathsf{witness}_{\mathsf{agg}}
=
\{(\mathsf{nonce}_i,k_i,\mathsf{tag}_i,d_i,b_i)\}_{i=1}^{m},
}
\]
{where each tuple corresponds to one user and satisfies the RSA accumulator non-membership relation with respect to the same public accumulator ${\mathcal{A}'(Y)}$.}

{Finally, the user submits the aggregated commitment $C_{\mathsf{agg}}$ and aggregated proof $\pi_{\mathsf{agg}}$ to the service provider. The verification process performs a single batch verification:}

\[
{
\mathsf{VerResult}
=
\mathsf{ZKVerify}
\left(
pp,\,
\pi_{\mathsf{agg}},\,
\{\mathsf{epoch},{\mathcal{A}'(Y)},C_{\mathsf{agg}}\}
\right).
}
\]

{If $\mathsf{VerificationResult}=\top$, the verifier accepts the batch, which implies that all constituent users have valid non-membership proofs with respect to the current blocklist accumulator. In this way, batch verification is achieved through a single aggregated proof, thereby significantly reducing communication overhead and computational complexity.}

\subsubsection{Dynamically and Directly Updating Aggregated Proofs}

When the blocklist is updated, the accumulator value ${\mathcal{A}(Y)}$ changes accordingly. 
To avoid recomputing all non-membership witnesses from scratch, $\mathsf{ShadowBlock}$ adopts an incremental witness-update mechanism for RSA accumulators. 
The key idea is that each user locally updates the witness associated with its own hidden representative, and then refreshes the aggregated zero-knowledge proof using the updated witness set.

Let
\[
u_Y=\prod_{y\in Y}y,\qquad {\mathcal{A}(Y)}=g^{u_Y}\bmod N,
\]
where each $y\in Y$ is the prime representative of a blocklisted tag, obtained by a hash-to-prime function.
For a non-blocklisted user with hidden representative $y_i\notin Y$, its non-membership witness is a pair $(d_i,b_i)$ satisfying
\[
d_i^{\,y_i}\cdot {\mathcal{A}(Y)}^{\,b_i}\equiv g \pmod N.
\]
Equivalently, there exist integers $(a_i,b_i)$ such that
\[
a_i y_i + b_i u_Y = 1,\qquad d_i=g^{a_i}\bmod N.
\]

\noindent\textbf{Case 1: Adding a member to the blocklist}

Suppose a new blocklisted identifier $\mathsf{tag}_{\mathrm{add}}$ is inserted, and let
\[
y_{\mathrm{add}}=H_{\mathsf{prime}}(\mathsf{tag}_{\mathrm{add}}).
\]
Then the updated hash set is
\[
Y' = Y \cup \{y_{\mathrm{add}}\},
\]
and the service provider updates the RSA accumulator as
\[
{\mathcal{A}'(Y)}={\mathcal{A}(Y)}^{\,y_{\mathrm{add}}}\bmod N.
\]

For each user $i$ with hidden representative $y_i\notin Y'$, let $(d_i,b_i)$ be the old witness with respect to $Y$. 
To derive the new witness, the user computes integers $(\alpha_i,\beta_i)$ such that
\[
\alpha_i y_i + \beta_i y_{\mathrm{add}} = b_i.
\]
Such integers exist because $y_i$ and $y_{\mathrm{add}}$ are distinct prime representatives and hence coprime.

The updated witness is then computed as
\[
d_i' = d_i\cdot {\mathcal{A}(Y)}^{\alpha_i}\bmod N,\qquad b_i'=\beta_i.
\]
Indeed, we have
\begin{align*}
(d_i')^{y_i}\cdot \mathcal{A}(X')^{b_i'}
&= \left(d_i\cdot {\mathcal{A}(Y)}^{\alpha_i}\right)^{y_i}
\cdot \left({\mathcal{A}(Y)}^{x_{\mathrm{add}}}\right)^{\beta_i} \\
&\equiv d_i^{y_i}\cdot {\mathcal{A}(Y)}^{\alpha_i y_i+\beta_i x_{\mathrm{add}}} \\
&\equiv d_i^{y_i}\cdot {\mathcal{A}(Y)}^{b_i} \\
&\equiv g \pmod N,
\end{align*}
which shows that $(d_i',b_i')$ is a valid non-membership witness for $y_i$ with respect to the updated set $Y'$.

After all affected users update their local witnesses, the corresponding aggregated proof is refreshed using the updated witness set. 
Importantly, this update does not require recomputing witnesses from the entire blocklist; it only requires a constant number of modular operations and one extended Euclidean computation per user.

\noindent\textbf{Case 2: Removing a member from the blocklist}

Suppose an existing blocklisted identifier $\mathsf{tag}_{\mathrm{del}}$ is removed, and let
\[
y_{\mathrm{del}}=H_{\mathsf{prime}}(\mathsf{tag}_{\mathrm{del}}).
\]
Then the updated hash set is
\[
Y'' = Y \setminus \{ y_{\mathrm{del}} \}.
\]
Let
\[
u_Y = u_{Y''}\cdot y_{\mathrm{del}}.
\]
The service provider updates the accumulator from ${\mathcal{A}(Y)}$ to ${\mathcal{A}''(Y)}$, where
\[
{\mathcal{A}''(Y)} = g^{u_{Y''}}\bmod N.
\]
Operationally, the update can be performed efficiently by the service provider using the trapdoor information of the RSA modulus, while users only need the new public accumulator value ${\mathcal{A}''(Y)}$.

For each user $i$ with hidden representative $y_i\notin Y''$, let $(d_i,b_i)$ be the old witness with respect to $Y$, namely
\[
d_i^{\,y_i}\cdot {\mathcal{A}(Y)}^{\,b_i}\equiv g \pmod N.
\]
Since
\[
u_Y = u_{Y''}\cdot y_{\mathrm{del}},
\]
the Bézout relation
\[
a_i y_i + b_i u_Y = 1
\]
can be rewritten as
\[
a_i y_i + (b_i y_{\mathrm{del}})u_{Y''}=1.
\]
Therefore, the updated witness with respect to $Y''$ is simply
\[
d_i'' = d_i,\qquad b_i'' = b_i\cdot y_{\mathrm{del}}.
\]
Its correctness follows from
\begin{align*}
(d_i'')^{y_i}\cdot {\mathcal{A}''(Y)}^{b_i''}
&= d_i^{y_i}\cdot {\mathcal{A}''(Y)}^{b_i x_{\mathrm{del}}} \\
&= d_i^{y_i}\cdot {\mathcal{A}(Y)}^{b_i} \\
&\equiv g \pmod N.
\end{align*}

Hence, under deletions, witness updates are even simpler: the user only scales the coefficient $b_i$ by the removed representative $y_{\mathrm{del}}$, while keeping $d_i$ unchanged.

\smallskip
After the local witness-update phase, the user recomputes the aggregation digest and generates a fresh aggregated zero-knowledge proof for the new public accumulator state. 
Compared with rebuilding all non-membership witnesses from scratch, this approach only updates the existing witnesses incrementally and therefore substantially reduces the cost of dynamic blocklist maintenance.

\subsubsection{Verification of Aggregated Proof}

Assume that a batch of $m$ users jointly submit an aggregated authentication proof. 
For each user $i\in\{1,\ldots,m\}$, let
\[
C_i = H\!\left(\mathsf{epoch}\,\|\,{\mathcal{A}(Y)}\,\|\,d_i\,\|\,b_i\right)
\]
be the commitment digest of the $i$-th updated witness, and let the aggregation digest be
\[
C_{\mathsf{agg}} = H(C_1\|\cdots\|C_m).
\]

The prover then generates a single aggregated zero-knowledge proof $\pi_{\mathsf{agg}}$ for the statement that, for all $i\in\{1,\ldots,m\}$:

\begin{enumerate}
    \item the hidden tag representative $y_i$ is correctly derived from the user's secret identity material and session nonce, i.e.,
    \[
    y_i = H_{\mathsf{prime}}(\mathsf{PRF}_{k_i}(\mathsf{nonce}_i));
    \]
    \item $y_i$ is a valid non-member of the current blocklist accumulator, i.e.,
    \[
    d_i^{\,y_i}\cdot {\mathcal{A}(Y)}^{\,b_i}\equiv g \pmod N;
    \]
    \item each $C_i$ and the final digest $C_{\mathsf{agg}}$ are computed correctly.
\end{enumerate}

Accordingly, the service provider verifies the batch by checking one aggregated zero-knowledge proof:
\[
\mathsf{VerResult}
=
\mathsf{ZKVerify}\!\left(
pp,\,
\pi_{\mathsf{agg}},\,
\{\mathsf{epoch},{\mathcal{A}(Y)},C_{\mathsf{agg}}\}
\right).
\]

If $\mathsf{VerificationResult}=\top$, the verifier accepts the whole batch, which means that all constituent users have proved, in zero knowledge, that their hidden representatives are not accumulated in the current blocklist.
Otherwise, the batch is rejected.

This design shifts batch efficiency to the aggregation layer: the verifier performs only one proof verification for the entire batch, while the prover updates each constituent RSA non-membership witness incrementally under additions and removals.
Therefore, $\mathsf{ShadowBlock}$ achieves efficient dynamic maintenance together with succinct batch verification.

\section{Security Analysis}
\label{sec:7}

\subsection{Formal Definitions}

{
We first introduce additional formal security definitions required for the aggregated proof system, including soundness and aggregated zero-knowledge preservation.
}

\begin{definition}[Blocklistability]
Let $\mathsf{Exp}^{\mathsf{BL}}_{\mathcal{A}}(\lambda)$ be the following experiment:

\noindent 1. Challenger runs $\mathsf{Setup}(1^\lambda)$ and initializes $X, Y$, and accumulator ${\mathcal{A}(Y)}$. \\
2. Adversary $\mathcal{A}$ may query registrations, updates, and the accumulator ${\mathcal{A}(Y)}$. \\
3. $\mathcal{A}$ outputs $(\mathsf{nonce}^*, \mathsf{tag}^*, \pi^*)$. \\
4. Challenger computes $y^* = H_{\mathsf{prime}}(\mathsf{tag}^*)$ and verifies:
\[
b = \mathsf{ZKVerify}(pp,\pi^*,({\mathcal{A}(Y)},N,g,\mathsf{nonce}^*)).
\]
5. The adversary wins if:
\[
(b=1 \wedge y^* \in Y) \vee (b=0 \wedge y^* \notin Y).
\]

We define:
\[
\mathsf{Adv}^{\mathsf{BL}}_{\mathcal{A}}(\lambda)
= \Pr[\mathsf{Exp}^{\mathsf{BL}}_{\mathcal{A}}(\lambda)=1].
\]

\end{definition}

\begin{definition}[Anonymity]
The anonymity experiment is defined as in the original manuscript.
{
Formally, it follows the standard indistinguishability game where the adversary cannot distinguish which honest user generated the proof due to PRF pseudorandomness and zkSNARK zero-knowledge property.
}
\end{definition}

\begin{definition}[Non-frameability]
The non-frameability experiment is defined as in the original manuscript.
{
The adversary is allowed adaptive queries to honest users and succeeds only if it can produce a valid accepting proof that falsely implicates an honest user.
}
\end{definition}

\begin{definition}[Soundness of Aggregated Proof System]
The aggregated proof system is sound if for any PPT adversary $\mathcal{A}$,
it is infeasible to generate an accepting aggregated proof $\pi_{\mathsf{agg}}$
such that at least one underlying statement is false, i.e., there exists $y_j \in Y$
such that $\mathcal{A}$ convinces the verifier that $y_j \notin Y$.
\end{definition}

\begin{definition}[Aggregated Zero-Knowledge Preservation]
The aggregated proof system preserves zero-knowledge if for any PPT adversary $\mathcal{A}$,
there exists a simulator $\mathcal{S}$ such that:
\[
\mathsf{View}_{\mathcal{A}}^{\text{real}} \approx \mathsf{View}_{\mathcal{A}}^{\text{sim}},
\]
where the view includes $\pi_{\mathsf{agg}}$, $C_{\mathsf{agg}}$, and public inputs
${(\mathcal{A}(Y), N, g, \mathsf{epoch})}$, and no witness information is leaked.
\end{definition}

\subsection{Security Analysis}

\begin{theorem}[Blocklistability]
Let $\mathcal{A}$ be a PPT adversary. If RSA accumulator is sound,
Groth16 is knowledge-sound, and PRF is secure, then:
\[
\mathsf{Adv}^{\mathsf{BL}}_{\mathcal{A}}(\lambda) \le \mathsf{negl}(\lambda).
\]
\end{theorem}

\begin{proof}
Assume adversary wins with non-negligible probability.

By Groth16 knowledge-soundness, there exists extractor $\mathcal{E}$ producing witness:
\[
(k,\mathsf{nonce},\mathsf{tag},y,d,b)
\]
such that:
\[
d^{y}\cdot {\mathcal{A}(Y)}^b \equiv g \pmod N.
\]

If $y \in Y$, this contradicts RSA accumulator soundness.

If $y \notin Y$ but accepted/rejected inconsistently, contradicts completeness.

Thus contradiction holds.
\end{proof}

\begin{theorem}[Anonymity]
{
If PRF is pseudorandom, Groth16 is zero-knowledge, and commitments are hiding,
then no PPT adversary can distinguish proofs generated by different users.
}
\[
\mathsf{Adv}^{\mathsf{Anon}}_{\mathcal{A}}(\lambda) \le \mathsf{negl}(\lambda).
\]
\end{theorem}

\begin{proof}
{
The adversary's view consists of PRF outputs, zkSNARK proofs, and commitments.
PRF ensures indistinguishability of tags, zkSNARK ensures witness hiding,
and commitment hiding ensures no leakage of internal state.
Thus distinguishing advantage is negligible.
}
\end{proof}

\begin{theorem}[Non-frameability]
{
If PRF is secure, commitments are binding, zkSNARK is knowledge-sound,
and RSA accumulator is sound, then non-frameability holds.
}
\[
\mathsf{Adv}^{\mathsf{NF}}_{\mathcal{A}}(\lambda) \le \mathsf{negl}(\lambda).
\]
\end{theorem}

\begin{proof}
{
Any successful framing attack implies at least one of:
(1) PRF forgery,
(2) zkSNARK forgery,
(3) accumulator soundness violation,
(4) commitment binding violation.
Each contradicts standard cryptographic assumptions.
}
\end{proof}

\begin{theorem}[Aggregated Soundness]
If zkSNARK is knowledge-sound and RSA accumulator is sound, then:
{
the aggregated proof system is sound.
}
\end{theorem}

\begin{proof}
Assume adversary produces valid aggregated proof but some $y_j \in Y$.

Extractor obtains witness:
\[
(d_j,b_j,y_j)
\]
satisfying:
\[
(d_j)^{y_j}\cdot {\mathcal{A}(Y)}^{b_j} \equiv g.
\]

This contradicts RSA accumulator soundness.
\end{proof}

\begin{theorem}[Aggregated Zero-Knowledge Preservation]
If Groth16 is zero-knowledge and commitments are hiding, then:
{
aggregated proof system preserves zero-knowledge.
}
\end{theorem}

\begin{proof}
{
Simulator generates each sub-proof using Groth16 simulator and combines them.
Commitments are simulated using hiding property.
Thus real and simulated views are indistinguishable.
}
\end{proof}

\section{Implementation and Evaluation}
\label{sec:8}
In this section, we present our prototype system implementation and the evaluation results. 

\subsection{Prototype Implementation}

Our implementation is single-threaded, and all our experiments were performed on an Intel® Core™ i5-8259U CPU @ 2.30GHz × 4 for the client side, and Intel Xeon Platinum 8480C with 530GB DDR5 4800MHz ECC RAM for the server side. Unless stated otherwise, we perform 3 runs of each experiment and report the average. We implement the membership proof aggregation algorithm from Boneh et al. \cite{7} to aggregate a set of membership proofs in RSA accumulators.

\subsection{Comparison of Works}

We conduct extensive experiments to evaluate the practicability, functionality, and efficiency of $\mathsf{ShadowBlock}$. Specifically, we compare $\mathsf{ShadowBlock}$ with BLAC \cite{blac}, SNARKblock \cite{SNARKBlock}, MaskAuct \cite{MaskAuct}, and ALPACA \cite{ALPACA}. BLAC \cite{blac} is a very classic anonymous blocklisting scheme, which is the first scheme for an anonymous blocklist based on the zero-knowledge proof. SNARKblock \cite{SNARKBlock} is pioneering in its approach; it significantly enhances the efficiency of blocklisting and improves its practicality. MaskAuct \cite{MaskAuct} is the latest application improvement of an anonymous blocklist in the field of auctions. ALPACA \cite{ALPACA} is the most advanced anonymous blocklist scheme that we are aware of. This comprehensive evaluation and comparison demonstrate that the $\mathsf{ShadowBlock}$ protocol has better dynamism and efficiency than the most advanced methods.

\begin{figure*}[htbp]
  \centering
  \subfloat[Running time for proof generation.]{
  \begin{minipage}[t]{0.3\textwidth}
    \centering
    \includegraphics[width=6cm]{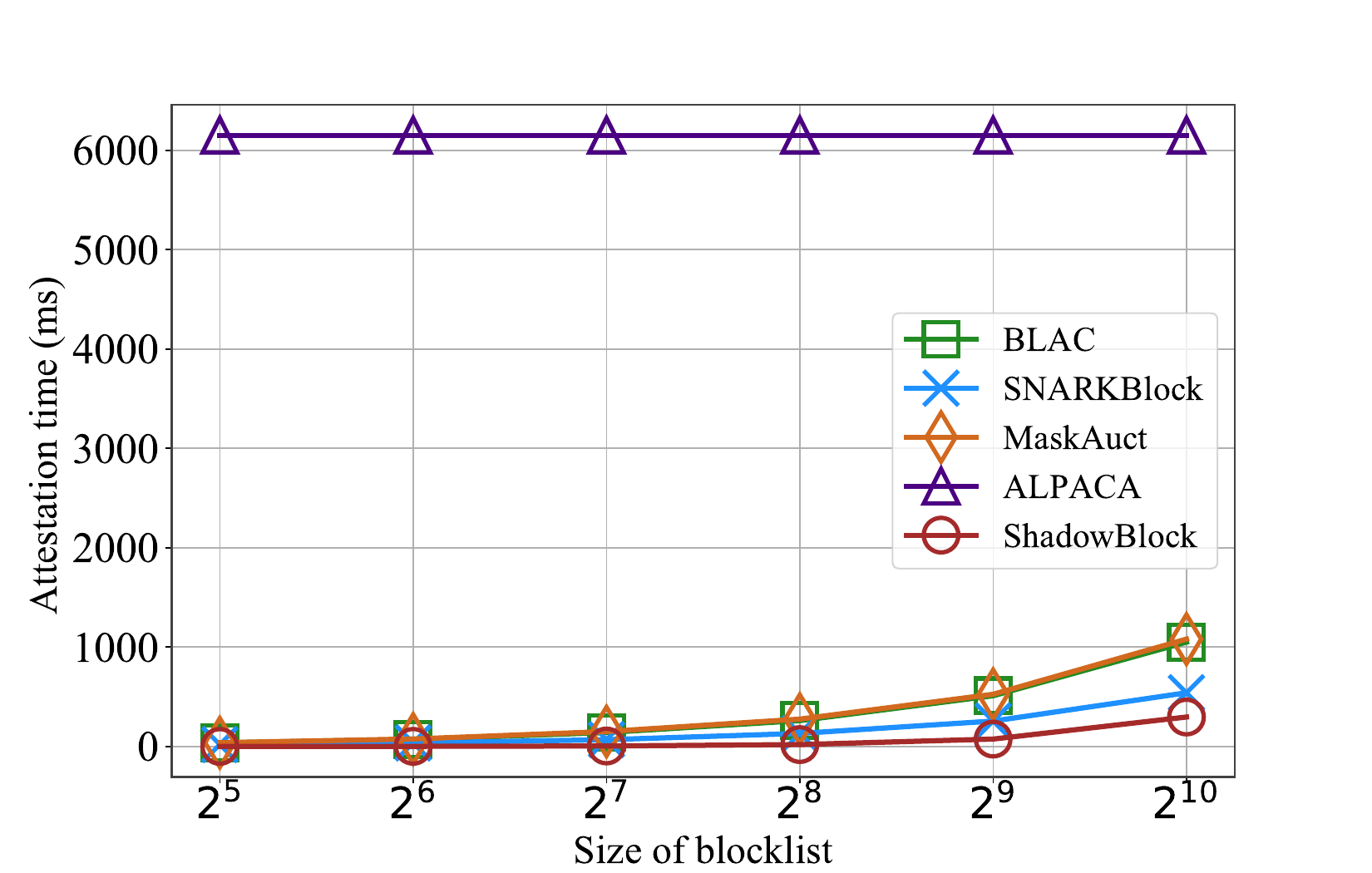}
  \end{minipage}}\hspace{0.4cm}
  \subfloat[Running time for verification.]{
  \begin{minipage}[t]{0.3\textwidth}
    \centering
    \includegraphics[width=6cm]{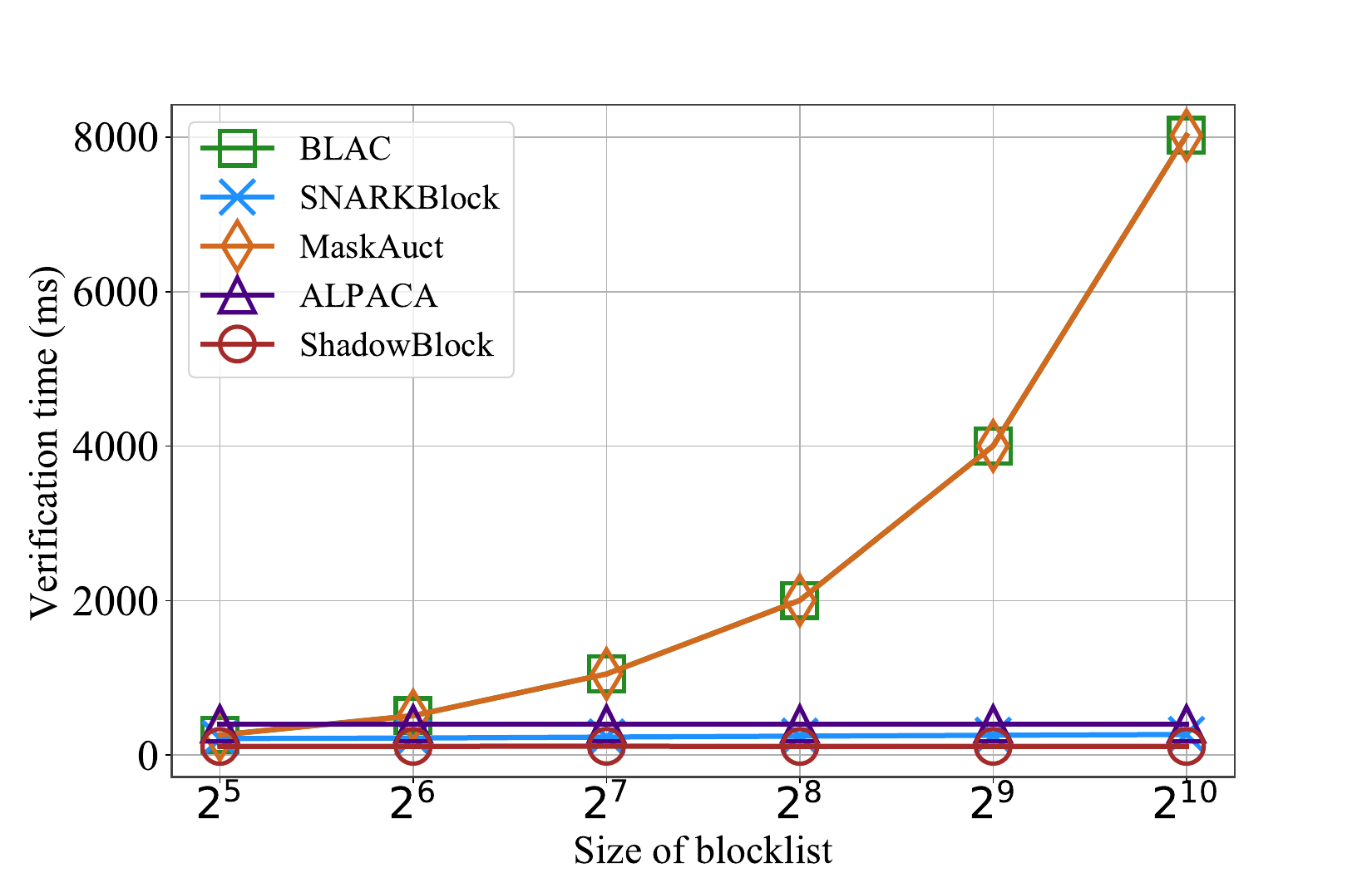}
  \end{minipage}}\hspace{0.5cm}
  \subfloat[Running time for verification (exclude BLAC\cite{blac} and MaskAuct\cite{MaskAuct}).]{  
    \begin{minipage}[t]{0.3\textwidth}
    \centering
    \includegraphics[width=6cm]{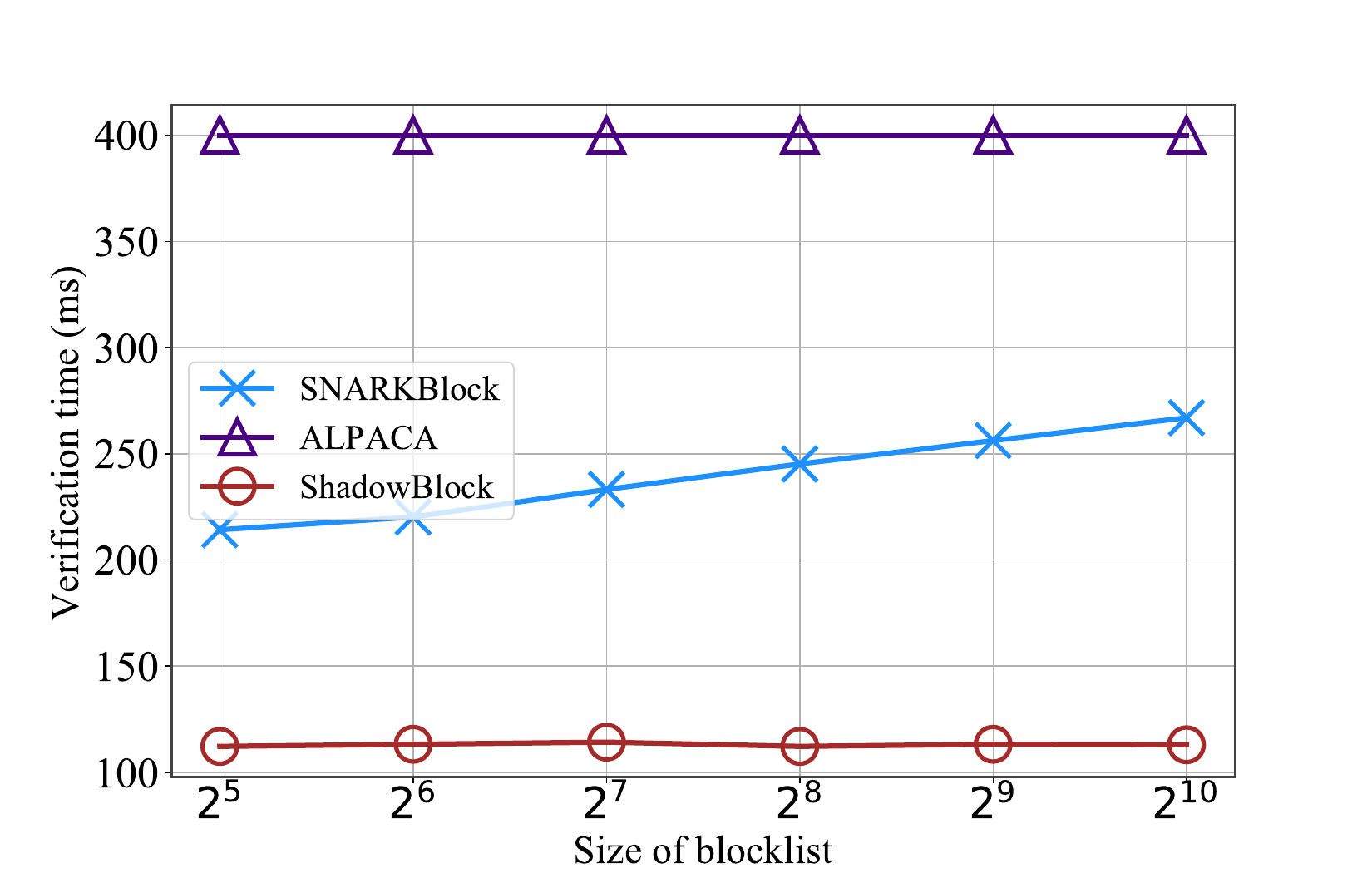}
  \end{minipage}}\hspace{0.38cm}

  \caption{{{Running time comparison}}}
  \label{fig:running}

\end{figure*}

\begin{figure*}[htbp]
  \centering
  \subfloat[Proof size.]{
  \begin{minipage}[t]{0.3\textwidth}
    \centering
    \includegraphics[width=6cm]{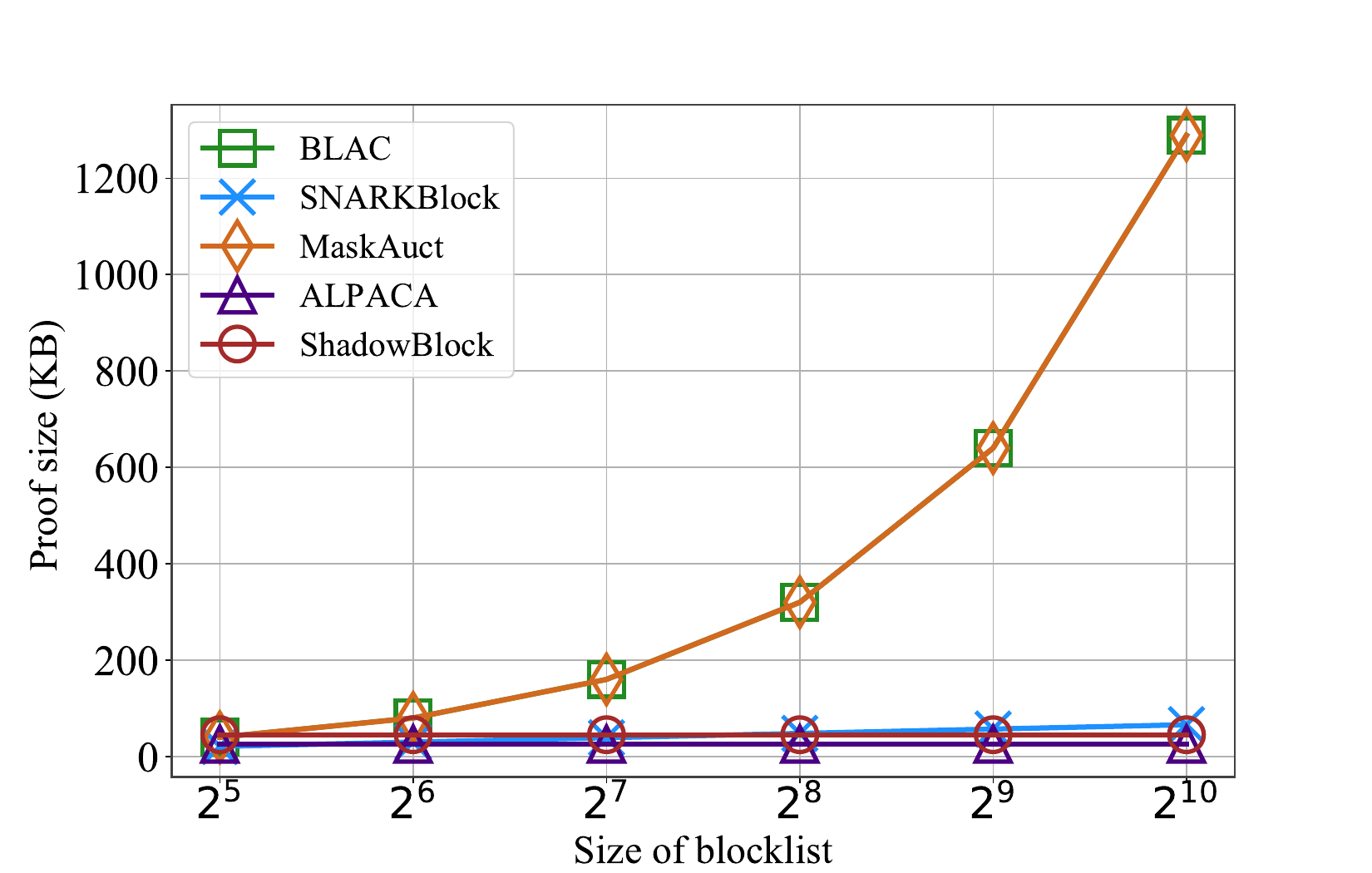}
  \end{minipage}}\hspace{0.4cm}
  \subfloat[Proof size (exclude BLAC\cite{blac} and MaskAuct\cite{MaskAuct}).]{
  \begin{minipage}[t]{0.3\textwidth}
    \centering
    \includegraphics[width=6cm]{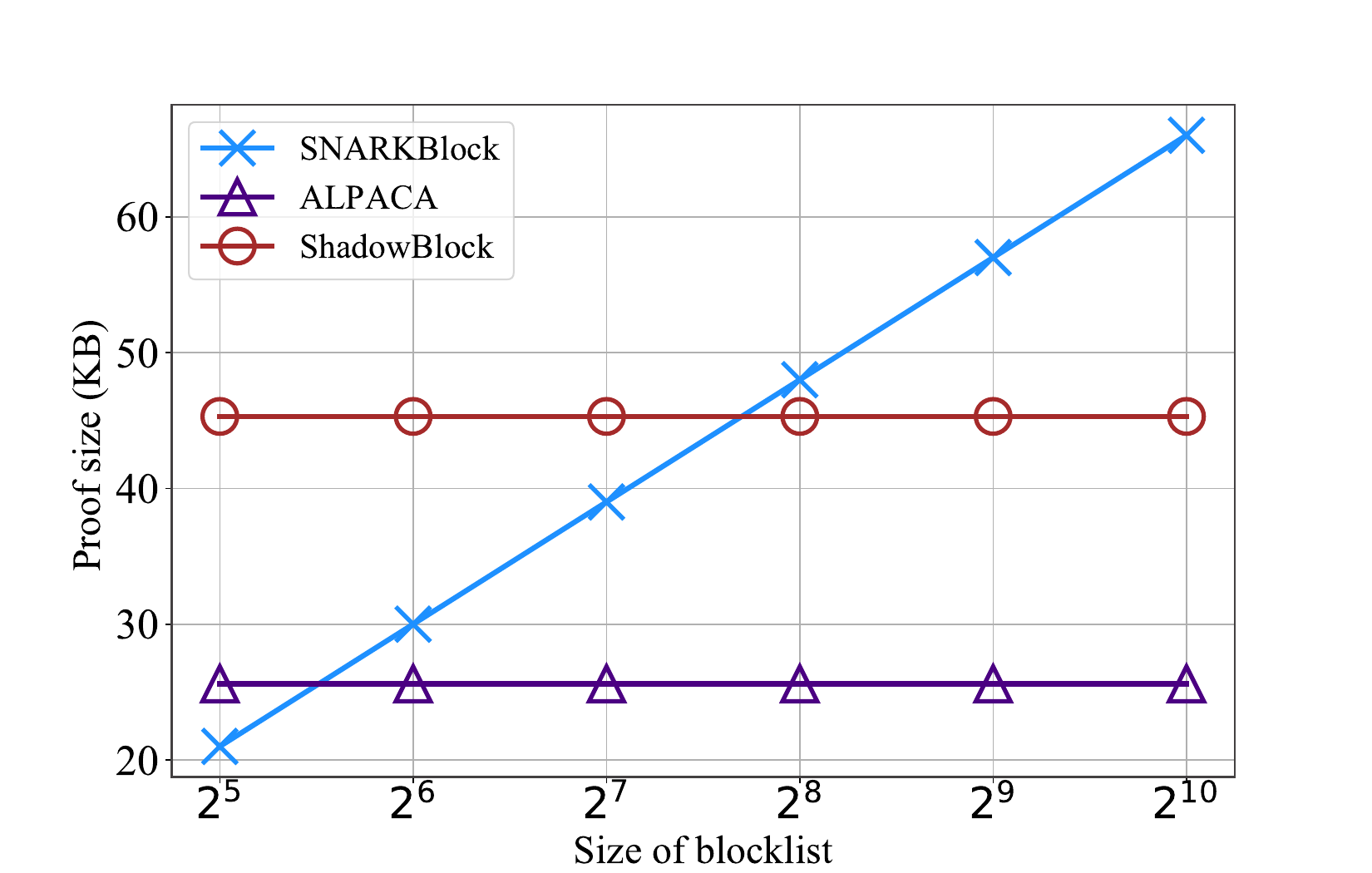}
  \end{minipage}}\hspace{0.5cm}
  \subfloat[Blocklist storage.]{  
    \begin{minipage}[t]{0.3\textwidth}
    \centering
    \includegraphics[width=6cm]{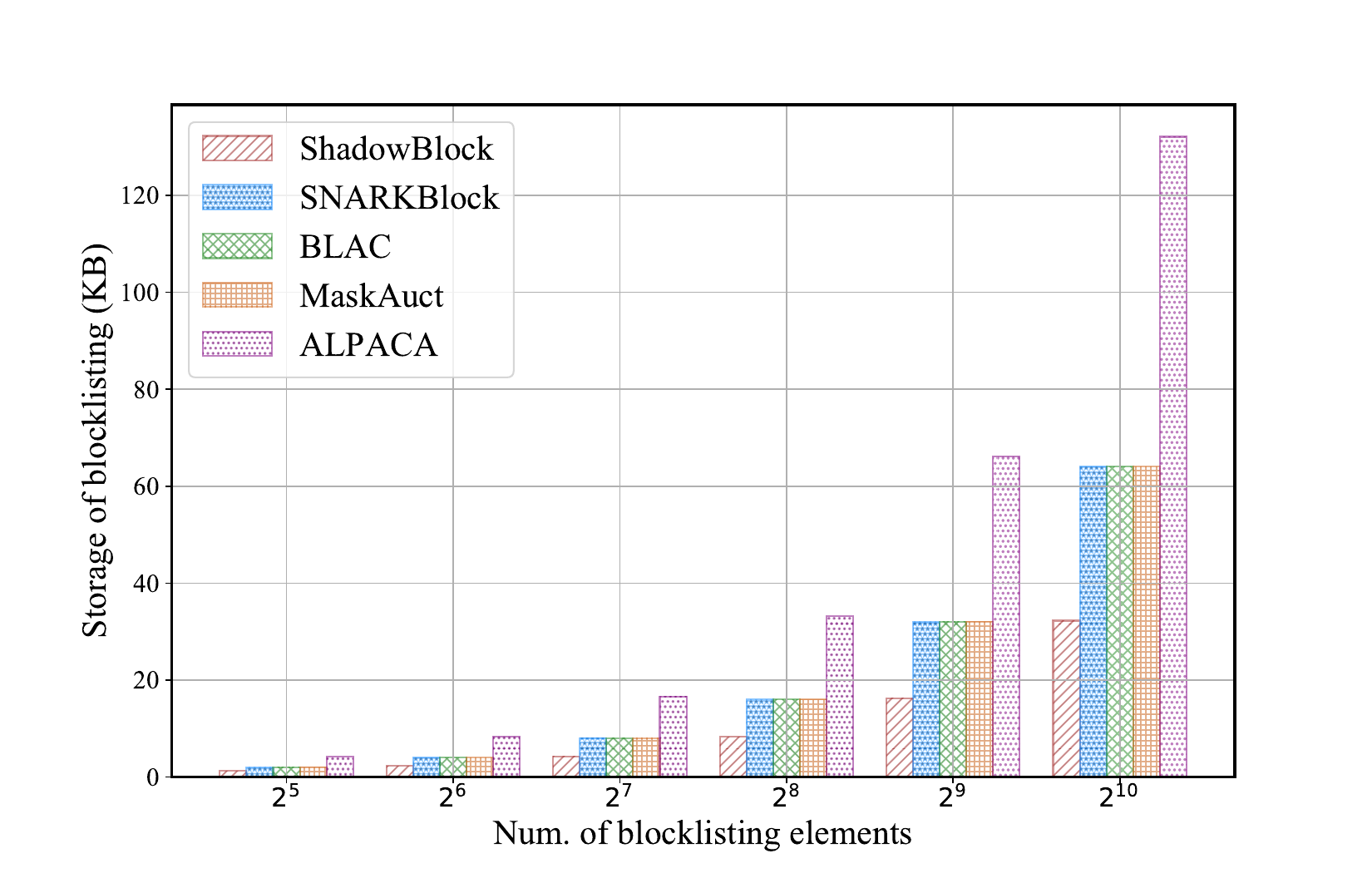}
  \end{minipage}}\hspace{0.38cm}
  \caption{{{Size comparsion}}}
  \label{fig:size}
\end{figure*}

\begin{table*}[htbp]
\centering
\caption{Client-Side Time Overhead for Blocklist Updates}
\label{tab:overhead_comparison}
\begin{tabular}{@{}l|c|cccc@{}}
\toprule
\multirow{2}{*}{\textbf{Functionality}} & \multirow{2}{*}{\textbf{Number (n)}} & \multirow{2}{*}{\textbf{$\mathsf{ShadowBlock}$}} & \multirow{2}{*}{\textbf{ALPACA}} & \multirow{2}{*}{\textbf{SNARKBlock}} & \textbf{BLAC} \\
 &  &  &  &  & \& \textbf{MaskAuct} \\ \midrule

 \multirow{5}{*}{\textbf{Addition}} & 2 & 0.30 s & 7.44 s & 0.54 s & 1.07 s \\
& 6 & 0.31 s & 9.92 s & 0.55 s & 1.06 s \\
& 10 & 0.32 s & 12.58 s & 0.57 s & 1.09 s \\
& 14 & 0.33 s & 15.04 s & 0.54 s & 1.07 s \\
& 18 & 0.38 s & 17.51 s & 1.12 s\textsuperscript{*} & 1.05 s \\ \midrule
\multirow{5}{*}{\textbf{Removal}} & 2 & 0.31 s & \textbf{0 s} / 7.46 s\textsuperscript{\dag} & 0.54 s & 1.05 s \\
& 6 & 0.33 s & \textbf{0 s} / 9.94 s\textsuperscript{\dag} & 0.56 s & 1.09 s \\
& 10 & 0.34 s & \textbf{0 s} / 12.52 s\textsuperscript{\dag} & 0.54 s & 1.04 s \\
& 14 & 0.36 s & \textbf{0 s} / 15.01 s\textsuperscript{\dag} & 0.53 s & 1.06 s \\
& 18 & 0.37 s & \textbf{0 s} / 17.58 s\textsuperscript{\dag} & 0.55 s & 1.10 s \\ \bottomrule

\multicolumn{6}{l}{\footnotesize \textsuperscript{*} Indicates Buffer Overflow.} \\
\multicolumn{6}{l}{\footnotesize \textsuperscript{\dag} The second value applies only to the specific user being unblocked.}
\end{tabular}
\end{table*}

\subsection{Performance Evaluation}

This set of figures presents a performance comparison of $\mathsf{ShadowBlock}$ with BLAC \cite{blac}, SNARKBlock \cite{SNARKBlock}, MaskAuct \cite{MaskAuct}, and ALPACA \cite{ALPACA} protocols in various scenarios. It should be noted that for the blocklisting solution using the chunks and buffer structure, we set the size of each chunk to $16 = 2^4$. The experimental results are presented in Figure \ref{fig:running}, Figure \ref{fig:size}, and Table \ref{tab:overhead_comparison}. We provide a detailed analysis of these results across five key metrics: proof generation time, verification time, proof size, blocklist storage overhead, and update cost.

\textbf{Proof Generation:} We experimented to compare the time required for generating non-blocklist non-membership proofs across different blocklist sizes. As illustrated in Figure \ref{fig:running}(a), the five solutions show distinct performance divergence as the blocklist size increases. $\mathsf{ShadowBlock}$ consistently maintains millisecond-level latency, rising smoothly, thus demonstrating both optimal performance and excellent scalability. SNARKBlock exhibits an approximately linear increase in latency with the blocklist size, reaching about 540 ms at the maximum scale. By contrast, BLAC and MaskAuct exhibit a steeper growth rate, with their latencies exceeding 1s when the blocklist size reaches $2^{10}$. In comparison, ALPACA's proof generation time is independent of the blocklist size but incurs a substantial constant overhead of 6.15s due to its IVC-based design, rendering it unsuitable for real-time authentication scenarios. The attestation time of BLAC is significantly longer than that of the other schemes because it needs to prove and generate all blocklist elements each time a non-blocklist membership proof is produced. SNARKBlock and MaskAuct achieve comparable attestation times, since MaskAuct is inspired by SNARKBlock and adopts the same chunk-and-buffer blocklist structure, which partitions the blocklist into a combination of multiple chunks and a buffer. Non-blocklist membership proofs for chunk elements can be reused, thereby significantly reducing proof generation time only the non-membership proofs for buffer elements need to be generated on demand. However, the integration of MaskAuct with Dynamic Group Signature (DGS) \cite{DGS} introduces additional computational overhead; consequently, MaskAuct has a shorter attestation time than SNARKBlock. Overall, $\mathsf{ShadowBlock}$ clearly demonstrates superior efficiency for dynamic, low-latency application environments.

\textbf{Verification:} In the verification-time experiment, as shown in Figure \ref{fig:running}(b) \& (c), $\mathsf{ShadowBlock}$ maintains a nearly constant and minimal cost across all blocklist sizes, owing to its aggregated proof design, which keeps verification complexity independent of the blocklist. SNARKBlock shows moderate growth as the blocklist expands because its chunk–buffer structures introduce additional verification work. BLAC and MaskAuct perform the worst, with verification time rising rapidly to several seconds due to their need to process proofs proportional to the full blocklist. ALPACA achieves size-independent verification but incurs a much higher constant cost of about 400 ms because of its IVC-based recursive proof system. Overall, $\mathsf{ShadowBlock}$ delivers the fastest and most scalable verification performance among all schemes.

\textbf{Proof Size:} As shown in Figure \ref{fig:size}(a) \& (b), the curves for $\mathsf{ShadowBlock}$ and ALPACA are perfectly horizontal lines, indicating that the proof size of these two schemes is independent of the blocklist scale. This is because $\mathsf{ShadowBlock}$ leverages a dynamic accumulator and aggregate zero-knowledge proofs, so the length of the generated zero-knowledge proof does not increase with the number of elements, with only a slightly larger constant factor. In contrast, ALPACA relies on iterative verification computation (IVC); each recursive update maintains a fixed-size recursive SNARK, and only this single recursive proof needs to be checked during final verification, thus resulting in a strictly constant proof size. By comparison, the proof sizes of SNARKBlock and MaskAuct exhibit a slow upward trend as the blocklist scale increases. This stems from their chunk-and-buffer structure: the blocklist is partitioned into multiple chunks, each with its own commitment or Merkle root, and the proof needs to carry public inputs, paths, or auxiliary information associated with several chunks. As the blocklist expands, the number of chunks increases accordingly, leading to a rise in the structural data to be included in the proof, hence an approximately linear growth pattern. BLAC and MaskAuct perform the worst in this regard, with their proof size growing in an almost exponential manner. The fundamental reason lies in its design, which does not adopt accumulators, chunk-based segmentation, or aggregation techniques. The proof structure has an essentially linear dependence on the blocklist; a larger blocklist requires more content to be described in the circuit and witness, leading to a rapid expansion of the proof length. Overall, $\mathsf{ShadowBlock}$ clearly demonstrates superior efficiency in terms of both latency and proof size for dynamic, low-latency application environments.

\textbf{Blocklist storage:} In this storage-overhead experiment, shown in Figure \ref{fig:size}(c). This figure illustrates the variation in storage space required by each scheme as the blocklist size scales from $2^5$ to $2^{10}$. It can be observed that the curves for SNARKBlock, BLAC, and MaskAuct overlap completely, indicating that these three schemes are essentially identical in their storage architecture: a full record (e.g., nonce and tag, or equivalent tuple) is stored for each blocklisted user. Consequently, doubling the blocklist size leads to an almost exact doubling of storage usage, representing a standard linear growth pattern. The curve for $\mathsf{ShadowBlock}$ exhibits the same linear shape but with significantly lower values (approximately half of the others at the same scale). This is because $\mathsf{ShadowBlock}$ compresses all tags into a fixed-size value using an accumulator, such that only all nonces plus a single accumulator need to be stored on-chain—rather than a separate tag for each user—resulting in lower overall storage consumption for the same number of elements. ALPACA incurs the highest overhead with linear growth but a steeper slope. This is due to the need to maintain additional recursive states beyond blocklist entries to enforce an append-only blocklist structure, which supports the IVC (Incremental Verifiable Computation) framework and subsequent updates. Each additional banned user thus increases the associated "administrative overhead," leading to a notably higher average storage cost per element compared to other schemes. Overall, all five schemes demonstrate consistent linear growth with increasing scale. However, $\mathsf{ShadowBlock}$ outperforms alternative designs in storage efficiency by compressing per-user information via an accumulator, while ALPACA incurs additional storage costs to enable its powerful recursive proof capabilities.

\textbf{Update:} For the experiment of updating the blocklist, the initial blocklist size is $2^{10}$ blocklist members, and the buffer setting of SNARKblock is an upper limit of 16 blocklist members. The results are shown in Table \ref{tab:overhead_comparison}.

\begin{itemize}
    \item For addition, $\mathsf{ShadowBlock}$ demonstrates an overwhelming performance advantage due to its advanced architectural design. Experimental data indicate that as the number of added members increases from 2 to 18, the total client-side time consumption for $\mathsf{ShadowBlock}$ remains consistently ultra-low, ranging between 0.30 s and 0.35 s. This efficiency stems from its algebraic update mechanism based on RSA accumulators: clients only need to perform millisecond-level parameter updates followed by the generation of an extremely lightweight zero-knowledge proof (baseline time of only 0.296s), thereby achieving sub-second response speeds. In contrast, SNARKBlock, while delivering excellent baseline performance (0.54 s) that outperforms traditional schemes like BLAC (1.05 s), suffers from stability issues due to its buffer mechanism. Performance is stable when the buffer is not full, but once the addition count (e.g., n=18) triggers a buffer overflow, the latency doubles to 1.10 s, revealing a risk of performance jitter. ALPACA, constrained by a high fixed proof overhead of 6.15 s and linear recursive folding costs, starts with a high latency of 7.4 s and rapidly climbs to 17.5 s as updates increase, making it unable to compete with $\mathsf{ShadowBlock}$ in terms of response speed.
    \item In the removal scenario, $\mathsf{ShadowBlock}$ reaffirms its position as the optimal solution. Benefiting from the mathematical symmetry of modular inverse operations, its deletion complexity is identical to that of addition, keeping the time cost locked between 0.30 s and 0.35 s. This is nearly 1.8 times faster than SNARKBlock and over 3 times faster than BLAC. This consistency ensures that all users experience extremely smooth performance regardless of network changes. Conversely, ALPACA exhibits extreme polarization: while the vast majority of unrelated users enjoy 0 s latency due to its logical deletion mechanism, the unblocked target user faces a severe penalty. To restore their identity, they must recompute all historical states since being unblocked, incurring a latency of 7.4 s to 17.5 s, effectively imposing the highest system delay on the user who needs service restoration the most. SNARKBlock maintains a respectable 0.54 s for removals, but it still lags behind $\mathsf{ShadowBlock}$'s sub-second performance. Furthermore, this operation forces all affected users to recompute proofs, creating a broader computational impact compared to $\mathsf{ShadowBlock}$'s lightweight updates. In summary, $\mathsf{ShadowBlock}$ is the only solution capable of providing robust, ultra-fast, and uniform experiences across all scenarios.
\end{itemize}

\section{Applications In Cross-chain}
\label{sec:9}

This section introduces an innovative cross-chain identity management \cite{cross-chain-identity} application built upon our $\mathsf{ShadowBlock}$ framework. We start by outlining the motivation, system architecture and detailing its workflow.

\subsection{Motivation}

{Cross-chain systems face distinct privacy and consistency challenges that do not exist in single-chain environments. When users interact across multiple chains via DIDs, credentials, or interoperable accounts, their activities become easily linkable. Using exposed identifiers or credentials for blocklisting allows moderation actions on one chain to reveal identity correlations across others, severely harming privacy. Meanwhile, cross-chain governance requires a globally consistent blocklist, creating a fundamental conflict between shared state coordination and user anonymity.}

{Cross-chain interaction \cite{cross-chain} further extends closed intra-chain identity systems into multi-chain environments, greatly raising the risk of cross-chain behavior tracing. Although DIDs support self-sovereign identity, their public nature makes on-chain enforcement actions universally visible if directly used for blocklisting, destroying anonymity. There are already blockchain payment solutions \cite{wang2025epass} with privacy protection on a single chain, but they are not applicable in a cross-chain environment. More seriously, in generative AI-driven mobile edge and satellite networks \cite{zhang3}, heterogeneous blockchains \cite{zhang1} leak metadata that enables cross-chain deanonymization attacks, where adversaries correlate events to reconstruct user profiles. Non-anonymous blocklisting thus turns cross-chain interoperability into a privacy leakage amplifier.}

{$\mathsf{ShadowBlock}$ effectively addresses these issues. It generates unlinkable pseudonymous tags via PRFs to avoid cross-chain identity correlation, employs compact accumulators for efficient shared blocklist maintenance, and supports dynamic updates for practical cross-chain moderation. Combined with zero-knowledge non-membership proofs, it expresses blocklisting only as mathematical set operations without exposing user identities, thus minimizing identity exposure and preserving privacy and untraceability in cross-chain ecosystems. Compared with single-chain deployment, the cross-chain setting introduces two additional issues. First, blocklist updates must be propagated across chains in a consistent manner. Second, proof validity must be tied to the freshness of the synchronized blocklist state. These requirements make dynamic anonymous blocklisting substantially more challenging in cross-chain systems than in conventional single-domain access control.}

\subsection{Cross-chain System Model}
\begin{figure}[htpb!]
	\centering
	\includegraphics[width=0.4\textwidth]{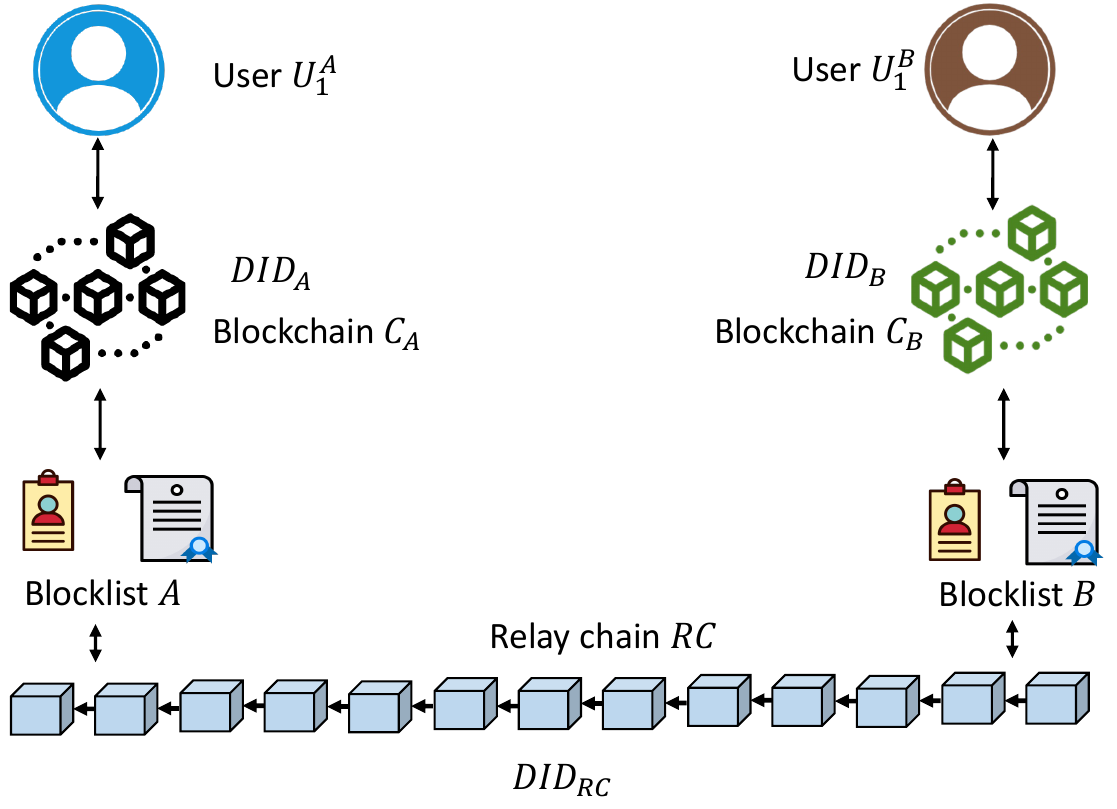}
	\caption{Cross-chain blocklist}\label{crosschain}
\end{figure}

Cross-chain platforms serve as critical infrastructures enabling interoperability among heterogeneous blockchain ecosystems. As illustrated in Fig.~\ref{crosschain}, blockchains $C_A$ and $C_B$ represent independent execution environments, while a relay chain $RC$ acts as the trust and messaging backbone connecting them. Each blockchain maintains its own Decentralized Identifier (DID) system following the W3C DID standard \cite{w3c-DID,w3c-VC}, whereas the relay chain holds a global blocklist accumulator shared across all chains.

$\mathsf{ShadowBlock}$ extends the conventional cross-chain DID architecture by incorporating an anonymous, accumulator-based blocklisting mechanism. Instead of exposing users’ real-world identities or DID attributes, users derive unlinkable pseudonymous tags using a PRF, $\mathsf{tag} = \mathsf{PRF}_{k}(\mathsf{nonce})$,
where $k$ is the user’s private key embedded in the DID credential, these PRF-derived tags are anonymized, session-independent identifiers that can be verified across chains without revealing any sensitive information. 

The relay chain maintains a global RSA accumulator ${\mathcal{A}(Y)}$ representing all banned pseudonymous identifiers {and prevents conflicting concurrent writes from producing inconsistent blocklist states across chains}. When a user seeks to access a service on a target blockchain, it must submit a non-membership zero-knowledge proof demonstrating that its tag is not included in the accumulator. This model enables all blockchains to enforce consistent global blocklisting decisions while ensuring strong anonymity and unlinkability.

\subsection{Workflow for Cross-Chain Identity Blocklist}

This section presents the complete workflow demonstrating how $\mathsf{ShadowBlock}$ enables privacy-preserving and verifiable cross-chain identity management. The workflow enhances the DID-based cross-chain authentication system by integrating accumulator-based anonymous blocklisting.

\subsubsection{Initialization}

The initialization procedure begins with setting up cryptographic parameters for RSA accumulators and zero-knowledge proofs. Each blockchain $C_i$ deploys a DID subsystem conforming to W3C standards, while the relay chain initializes a shared accumulator ${\mathcal{A}(Y)}$ and publishes the public parameters. The relay chain also establishes a consensus committee (e.g., PBFT-based) responsible for updating and verifying global blocklist states and synchronizing accumulator values across all chains.

\subsubsection{User Registration}

Users obtain decentralized identifiers from their local chain’s DID system. A DID document contains public keys and service endpoints, while private keys are controlled exclusively by the user. To interact across chains, a user applies a PRF on a randomly sampled nonce to derive an anonymous tag:
$\mathsf{tag} = \mathsf{PRF}_{k}(\mathsf{nonce})$,
where $\mathsf{nonce}$ is freshly sampled for each session to ensure uniqueness and prevent replay attacks. The corresponding accumulator representative is then computed as
$y = H_{\mathsf{prime}}(\mathsf{tag})$.
This tag ensures computational unlinkability across multiple chains.

\subsubsection{Blocklist Management}

Blocklist management is executed by the relay chain. When a violation is detected on any blockchain, the responsible chain computes the offender’s pseudonymous identifier $\mathsf{tag}$ and its corresponding representative $y = H_{\mathsf{prime}}(\mathsf{tag})$, and submits $y$ to the relay chain. The relay chain updates its accumulator as
\[
{\mathcal{A}'(Y)} = {\mathcal{A}(Y)}^{y} \bmod N,
\]
and broadcasts the updated accumulator to all chains.

Unblocking operations are supported through the accumulator’s dynamic removal capability. Let $y = H_{\mathsf{prime}}(\mathsf{tag})$ be the representative of the removed element. The updated accumulator is:
\[
{\mathcal{A}''(Y)} = g^{\prod_{y_i \in Y \setminus \{y\}} y_i} \bmod N,
\]
which can be efficiently computed by the relay chain using the trapdoor of the RSA accumulator. This ensures that the blocklist remains concise and prevents permanent accumulation of obsolete entries—an advantage over append-only designs such as ALPACA.

\subsubsection{User Authentication}

Before accessing a target chain, the user retrieves the current accumulator value ${\mathcal{A}(Y)}$ and generates a zero-knowledge non-membership proof:
\[
\pi = \mathsf{ZKProve}(pp,\text{witness}=(k,\mathsf{nonce},\mathsf{tag},y,d,b)).
\]
The proof asserts that the hidden representative $y = H_{\mathsf{prime}}(\mathsf{tag})$ is not included in the blocklist, i.e.,
\[
d^{\,y} \cdot {\mathcal{A}(Y)}^b \equiv g \pmod N,
\]
while revealing no information about the user’s identity, DID, or behavioral history. The target chain verifies the proof via:
\[
\mathsf{ZKVerify}(pp,\pi,({\mathcal{A}(Y)},N,g,\mathsf{nonce})) = 1.
\]
If verified, the user is granted access; otherwise, the request is rejected. This mechanism enables consistent and privacy-preserving blocklist enforcement across heterogeneous blockchain ecosystems.

\subsubsection{Cross-chain Optimization}
{To support high-frequency cross-chain operations, $\mathsf{ShadowBlock}$ employs three optimization techniques: \textbf{(1) Accumulator Synchronization:} A non-membership proof is valid only with respect to an explicit accumulator ${\mathcal{A}(Y)}$. The target chain accepts a proof only if ${\mathcal{A}(Y)}$ is the latest finalized relay-chain version already synchronized to the chain. \textbf{(2)Incremental Updates:}  When the blocklist changes, users update their non-membership proofs with minimal recomputation rather than regenerating the entire proof set. This dramatically reduces computational overhead. \textbf{(3)Aggregated Proofs:} Multiple non-membership proofs can be aggregated into a single succinct proof. The relay chain can batch-verify large batches of cross-chain authentication requests with near-constant cost, improving scalability in large cross-chain ecosystems.}

\section{Conclusion}
\label{sec:10}

$\mathsf{ShadowBlock}$ addresses the scalability and efficiency limitations of existing blocklisting systems by enabling dynamic and privacy-preserving anonymous blocklist management. It leverages pseudorandom functions and cryptographic accumulators to allow users to prove non-membership without revealing their identities. Zero-knowledge proofs ensure privacy, while proof aggregation reduces verification overhead. Meanwhile, accumulator-based constructions support efficient updates without full proof regeneration. Experimental results demonstrate that $\mathsf{ShadowBlock}$ outperforms prior works such as ALPACA and SNARKBlock in both efficiency and dynamism. The framework enables fast updates, compact storage, and strong security guarantees, making it well-suited for cross-chain environments.

\bibliographystyle{ieeetr}
\bibliography{reference}

\vfill

\end{document}